\documentclass[iop]{emulateapj}
\usepackage{epsfig,amsmath,natbib}
\usepackage{color}

\def\refjnl#1{{\rm#1}}
\def\araa{\refjnl{ARA\&A}}             
\def\apj{\refjnl{ApJ}}                 
\def\apjl{\refjnl{ApJ}}                
\def\apjs{\refjnl{ApJS}}               
\def\aap{\refjnl{A\&A}}                
\def\mnras{\refjnl{MNRAS}}             
\def\nat{\refjnl{Nature}}              
\def\jqsrt{\refjnl{J.~Quant.~Spec.~Radiat.~Transf.}} 

\def\refjnl#1{{\rm#1}}
\def\araa{\refjnl{ARA\&A}}             
\def\apj{\refjnl{ApJ}}                 
\def\apjl{\refjnl{ApJ}}                
\def\apjs{\refjnl{ApJS}}               
\def\aap{\refjnl{A\&A}}                
\def\mnras{\refjnl{MNRAS}}             
\def\nat{\refjnl{Nature}}              
\def\jqsrt{\refjnl{J.~Quant.~Spec.~Radiat.~Transf.}} 

\def\be{\begin{equation}} 
\def\ee{\end{equation}} 
\def\ba{\begin{eqnarray}} 
\def\ea{\end{eqnarray}}

\def\kms{\,{\rm {km\, s^{-1}}}} 
 
\def\kms{\,{\rm {km\, s^{-1}}}} 
\def\cc{\,{\rm {cm^{-3}}}} 
\def\ergs{\,{\rm erg\, {s^{-1}}}}

\def\BC{{\rm B}_{912}} 
\def\HI{\hbox{H~$\scriptstyle\rm I\ $}}

\def\erg{{\rm erg}} 
   
 \def\sec{{\rm s}} 
 
\def\gsim{\lower.5ex\hbox{\gtsima}} 
\def\lsim{\lower.5ex\hbox{\ltsima}} \def\gtsima{$\; \buildrel > \over 
\sim \;$} \def\ltsima{$\; \buildrel < \over \sim \;$} \def\prosima{$\; 
\buildrel \propto \over \sim \;$} \def\gsim{\lower.5ex\hbox{\gtsima}} 
\def\lsim{\lower.5ex\hbox{\ltsima}} 
\def\simgt{\lower.5ex\hbox{\gtsima}} 
\def\simlt{\lower.5ex\hbox{\ltsima}} 
\def\simpr{\lower.5ex\hbox{\prosima}}   
  
 \def\gtsima{$\; \buildrel > \over \sim \;$} 
\def\ltsima{$\; \buildrel < \over \sim \;$} 
\def\gsim{\lower.5ex\hbox{\gtsima}} 
\def\lsim{\lower.5ex\hbox{\ltsima}} 
\def\simgt{\lower.5ex\hbox{\gtsima}} 
\def\simlt{\lower.5ex\hbox{\ltsima}} 
\def\simpr{\lower.5ex\hbox{\prosima}}

\def\E3{{\cal E}_{\rm g}^{III}}

\def\r12{r_{1/2}} 
\def\x12{x_{1/2}} 
\def\v12{v_{1/2}}

\begin{document}

\title{On the Formation of Molecular Clumps in QSO Outflows} 
\author{Andrea Ferrara\altaffilmark{1}, Evan Scannapieco\altaffilmark{2}}
\altaffiltext{1}{Scuola Normale Superiore, Piazza dei Cavalieri 7, I-56126 Pisa, Italy}
\altaffiltext{2}{School of Earth and Space Exploration,  Arizona State University, P.O.  Box 871404, Tempe, AZ, 85287-1404}
 
 
 
\label{firstpage} 
\begin{abstract} 

We study the origin of the cold molecular clumps in quasar outflows, recently detected in CO and HCN emission. We first describe the physical properties of such radiation-driven outflows and show that a transition from a momentum- to an energy-driven flow must occur at a radial distance of $R \approx 0.25$ kpc.   During this transition, the shell of swept up material fragments due to Rayleigh-Taylor instabilities, but these clumps contain little mass and are likely to be rapidly ablated by the hot gas in which they are immersed.   We then explore an alternative scenario in which clumps form from thermal instabilities at $R \simgt 1$ kpc, possibly containing enough dust to catalyze molecule formation.   We investigate this processes with 3D two-fluid (gas+dust) numerical simulations of a  kpc$^3$ patch of the outflow, including atomic and dust cooling, thermal conduction, dust sputtering, and photoionization from the QSO radiation field.  In all cases, dust grains are rapidly destroyed in $\approx 10^4$ years; and while some cold clumps form at later times, they are present only as transient features, which disappear as cooling becomes more widespread. In fact, we only find a stable two-phase medium with dense clumps if we artificially enhance the QSO radiation field by a factor 100.  This result, together with the complete destruction of dust grains, renders the interpretation of molecular outflows a very challenging problem.          

\end{abstract}


\section{Introduction}
\label{Intro}

Cold, fast-moving molecular clumps have been detected at distances up to a few kpc from the central engines of a few dozen active galactic nucleii (AGN), mostly highly-obscured AGN in dusty star forming galaxies \citep{Feruglio15, Cicone2014, Veilleux13}. In the nearby QSO Markarian 231, for example, a $1000 M_\sun$ yr$^{-1}$ kpc-scale outflow with a radial velocity of $750-1000 \kms$ has been well detected by several groups  \citep{Feruglio10,Rupke11,Sturm2011}. Similarly, surveys of ultra-luminous infrared galaxies (ULIRGs) and quasar-hosts  suggest that the presence of an AGN can boost molecular outflow rates by large factors \citep{Sturm2011,Cicone2014}.

On the other hand, the origin of these molecular clumps is unclear. One possibility, in analogy with the prevalent picture for starburst driven winds, is that the clouds are driven out of the host galaxy by ram pressure acceleration \citep[e.g.][]{Veilleux2005}.   However this hypothesis runs into serious difficulties, both because: (i) shocks and conduction from the exterior medium tend to compress the clouds perpendicular to the direction of the flow, greatly reducing the momentum flux they receive; and (ii) instabilities and evaporation lead to rapid cloud disruption  \citep{Klein94,1994ApJ...434L..33M,2006A&A...457..545O,2008ApJ...678..274O,Scannapieco15,2016ApJ...822...31B}.   Together these effects imply that the lifetimes of the clumps are likely to be much shorter that the times required to accelerate them to the observed speeds.

A second possibility is that the clumps are formed from the cooling of very high temperature ($\approx 10^7$ K) shock-heated gas, already moving at high radial velocities.  In, fact  a similar picture has been suggested in the context of starburst-driven galaxy outflows \citep{1995ApJ...444..590W, 2000MNRAS.317..697E, 2003ApJ...590..791S, 2004ApJ...610..226S, 2007ApJ...658.1196T, 2011ApJ...740...75W, 2016MNRAS.455.1830T}, and ultra-luminous infrared galaxies hosting starbursts and AGN \citep{2015ApJ...803....6M}. However, in this case, one has not only to reconcile the clump formation process with the prevailing outflow scenario, but also explain how molecules like CO and HCN can form. These species require solid surfaces to trigger their production as gas phase production is likely to be highly inefficient in the presence of a strong UV background, and thus dust grains must be preserved throughout the entire heating/cooling cycle. This is not an easy requirement because grains are sputtered by nuclear and grain-grain collisions and thus rapidly destroyed in such hot environments.

The main motivation of this paper is to explore the possibility that the observed molecular clumps condense out of the quasar outflow material as a result of thermal instabilities occurring in the cooling gas.  Carrying out a detailed examination of the dynamical structure of QSO outflows, we conclude that the optimal location for the formation of dusty clumps is within the shocked gas during the energy-driven stage of the evolution.  We then carry out a set of detailed numerical simulations of clump formation at this stage, which include dust cooling and destruction,  electron thermal conduction, and atomic/ionic cooling and heating process that account for the QSO radiation field.    The inclusion of this extended number of physical processes allows us to draw a series of robust conclusions as to the feasibility of the formation molecular clumps though condensation in QSO driven outflows. 

The structure of this paper is as follows.  In \S2 we describe the physics of QSO driven outflows and the cooling of shocked material.  In \S3 we describe our initial conductions and simulation methods, and in \S4 we present our results.   We conclude in \S5  with a discussion of the implications of these results for clump formation in QSO outflows, and we include appendices that discuss the formation of multiphase media and self-shielding in further detail.  


\section{Physical Scenario}

\subsection{Overall Evolution}

\label{Met}
The model explored in this work is based on the following idealized assumptions. At $t=0,$ a radiation-pressure driven, relativistic wind ($v_w = \zeta c \simeq 0.1 c$), \citep{2003MNRAS.345..657K,2010MNRAS.402.1516K} emanates from the accretion disk powering a QSO. Such winds, with outflows rate of $\dot M_w \approx 1 M_\odot$yr$^{-1}$ have been clearly detected by nuclear X-ray observations \citep[e.g.][]{Murray1995,Crenshaw2003,Pounds2003}.  We assume that the wind material has a constant velocity $v_w$ and an outflow rate $\dot M_w \equiv dM_w/dt,$ such that the momentum rate is equal to
\be
\frac{dM_w}{dt} v_w \simeq  \frac{L_E}{c},
\label{eq01} 
\ee 
where $L_E$ is the classical Eddington luminosity of a black hole of mass $M_\bullet$,
\be
L_E = \frac{4\pi G M_\bullet c}{\kappa_{\rm es}} = 1.54\times 10^{38} \left(\frac{M_\bullet}{M_\odot}\right)  \erg \,\sec^{-1},
\label{eq02} 
\ee 
and $\kappa_{\rm es}=0.4 \, {\rm cm}^2$ g$^{-1}$ is the electron scattering opacity.
The corresponding kinetic energy input rate is 
\be
\frac{1}{2}\frac{dM_w}{dt} v_w^2 \simeq \frac{L_E^2}{2 \dot M_w c^2} \simeq \frac{\eta}{2} L_E \equiv L_w,
\label{eq03} 
\ee
where we have further assumed the outflow rate equals the Eddington rate, $\dot M_w \approx \dot M_E,$ and  $\eta \equiv L_E/ (\dot M_w c^2) \simeq 0.1$ is the canonical  radiative accretion efficiency \citep{Yu02}.
\begin{figure*}
\centerline{\includegraphics[width=170mm]{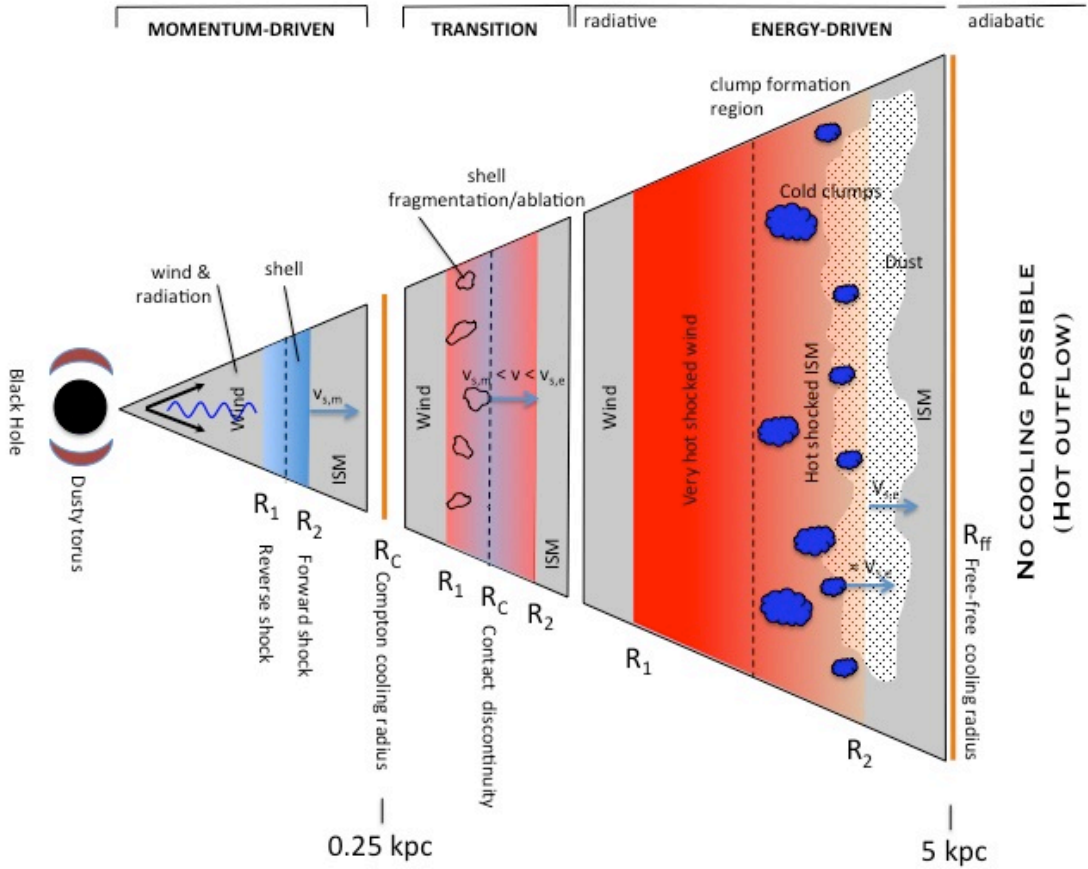}}
\caption{Schematic evolution of a typical QSO outflow showing the three different dynamical phases. The shell is initially driven by the wind momentum until it reaches the Compton cooling radius $R_c$. Beyond that point the shell is accelerated from the velocity $v_{s,m}$ to $v_{s,e}$, the velocity appropriate for the energy-driven phase. During such acceleration the shell is fragmented by Rayleigh-Taylor instabilities, the resulting clumps are rapidly ablated as they are entrained by the hot shock gas flowing past them, and the shell eventually dissolved. These short-lived clumps contain little mass and therefore cannot explain the observed molecular clumps. As the outflow enters the energy-driven phase, new, more massive, dusty clumps form from thermal instabilities occurring in the shocked ambient (ISM) gas within 5 kpc radius. These clumps are at rest with the hot gas and are not ablated.     
} 
\label{Fig1}
\end{figure*}

The wind then expands into the ambient interstellar medium (ISM) around the host galaxy, for which we assume an isothermal radial density profile, 
\be
\rho(R) = \frac {f_g \sigma^2}{2\pi G R^2}, 
\label{eq05} 
\ee
where $\sigma$ is the velocity dispersion of typical QSO host galaxy spheroids ($\sigma \simeq 200-300$ km s$^{-1},$) and $f_g$ is the gas fraction in the system. This number is bound from above by the cosmological value: $f_g < f_c = \Omega_b/\Omega_M$, but we allow for smaller values to account for the fact that some of the baryons are contained in stars. Eq.\ (\ref{eq05}) yields an enclosed gas mass within $R$ equal to
\be
M(<R) = \frac {2 f_g \sigma^2}{G} R. 
\label{eq06} 
\ee
The wind expansion drives a \textit{forward} spherical shock (see Fig.\ \ref{Fig1}) propagating in the ISM, whose radius we define as $R_1$, and a \textit{reverse} shock, located at a radius $R_2 < R_1$, propagating back into the wind. In between, a contact discontinuity exists at radius, $R_2 < R_c < R_1$, separating the shocked ISM and wind material. 

In the initial phases, wind parcels  passing through the reverse shock at $R_1$ are suddenly heated to a temperature $T \approx 3 \mu m_p v_w^2/16 k_B \approx 1.3 \times 10^{10}$ K,
where we have taken the mean molecular weight $\mu = 0.65$, appropriate for a fully ionized, solar metallicity plasma. Such plasma can in principle cool into a thin shell  located at $R_s \approx R_1$, via inverse Compton scattering with the QSO radiation field, again assumed to emit at $L_E$.\footnote{Note that for the same physical reason, there is also a second shell behind the forward shock, but as $R_1 \approx R_2$ in the momentum-driven phase, we treat them as a single entity.} In this case, the reverse shock is isothermal and the shell expansion is momentum-driven. As we will discuss in Sec. 3 later on, at larger ($\simgt$ kpc) radii Compton processes actually act as a heat source for the gas. 

Efficient cooling can be maintained only within a radius $R_C$, where the Compton cooling time is  
\be
t_C = \frac {4}{3}  \frac{c R_s^2}{GM_\bullet} \left(\frac{m_e}{m_p}\right)^2  \zeta^{-2}.
\label{eq04} 
\ee
For an isothermal density profile this occurs when the shell reaches a radius of 
\be
R_{C} = 0.25 \,\sigma_{200} \zeta^2_{-1}  M_{\bullet,8}^{1/2} \,\textrm{kpc},
\label{eq07} 
\ee
where $\zeta_{-1} \equiv v_w/c/0.1$.  Note that the outward velocity of the shell in the momentum-driven phase that is used in this expression is much less than $v_w$  (see derivation below).  

We pause here for a warning, as a problem might arise with the above scenario. The kinetic energy of the shock is transferred to protons, which therefore carry most of the thermal pressure, rather than electrons that mediate Compton cooling. Usually the equilibration of the two temperatures is rapidly ensured by Coulomb collisions and, under some conditions, plasma instabilities \citep{2012MNRAS.425..605F}. Therefore it is possible that the electron temperature, $T_e$ remains as low as expected from the minimal heating rate $T_e \approx (m_e/m_p)T_p$, where $T_p$ is the proton rate. In this case the gas cannot actually cool and the momentum-driven phase, along with the cold shell, cannot exist. On the other hand, a relativistic plasma ($T \simgt 10^{10}$ K) as that produced by the reverse shock may undergo pair production catastrophic cooling \citep{1987ApJ...313..689B}. This process, in principle, would make the presence of the QSO radiation field unnecessary to cool the gas and would force the outflow evolution to remain in a momentum-driven phase throughout the evolution.

Beyond $R_C$, the shocked wind material remains very hot and highly pressurized, so the outflowing motions are energy-driven, rather than momentum-driven. In this case, $1/2 \dot M_w v_w^2 \equiv L_w = (\eta/2) L_E$ is conserved. The pressure drives the forward shock into the ISM, and soon after, into the galactic halo. As the flow makes the transition to the energy-driven phase, the reverse shock detaches from the shell (i.e. $R_1 \ll R_s$). Below, we show that the shell likely fragments and dissolves in the hot gas during the transition. However, if it did survive, it would be driven by the pressure of the hot gas behind the reverse shock penetrating the wind. 

\cite{2003ApJ...596L..27K} notes that at the time of the transition to the energy-driven regime, the mass of the supermassive black hole attains a final value given by
\be
M_\sigma =  \frac {f_g \kappa_{es} \sigma^4}{\pi G^2}.
\label{eq08} 
\ee
The above scenario implies the existence of an initial momentum-driven shell followed by an energy-driven expansion.  Below we show that the gravitational deceleration due to the  galaxy and the central black hole can be safely neglected in our case, and thus the time evolution of the shell can be described by the classical  formalism  \cite[][eq.\ 21]{1977ApJ...218..377W} based on the solution of the momentum and energy equations: 
\be
\frac{d}{dt}\left[M(< R_s) \dot R_s\right] = \frac{L_E}{c} + 4\pi R_s^2 p, 
\label{eq09} 
\ee
\be
\frac{3}{2}\frac{d}{dt}\left(\frac{4\pi}{3} R_s^3 p\right) = L_w - 4\pi R_s^2 p \dot R_s - L_C(R_s),
\label{eq10} 
\ee
where $p$ is the pressure of the interior of the shell, which is due to the reverse shock propagating back to the center, and  $L_C$ is the energy loss rate of the gas due to Compton cooling. As discussed above in the initial momentum-driven phase, when $R_s \lesssim R_C,$ this term dominates the energy equation, and the pressure of the gas heated by the reverse shock is rapidly radiated away. As a result $p \ll L_E/4\pi R_s^2 c$, and the shell motion is fully described by eq.\ (\ref{eq09}) with $p \approx 0.$ Taking $M(<R)$ from eq.\ (\ref{eq06}) this gives
\be
\frac{2f_g \sigma^2}{G}\frac{d}{dt}(R_s\dot R_s) \simeq \frac{L_E}{c}, 
\label{eq11} 
\ee
which has the solution $R_{s,m} = v_{s,m} t$, where the shell velocity, $v_{s,m}$ in the momentum-driven phase is a constant value 
\be
v_{s,m} = \left(\frac{GL_E}{2 f_g \sigma^2 c}\right)^{1/2}.
\label{eq12} 
\ee

Let us now examine the other asymptotic behavior in which $L_E/c \ll 4\pi R_s^2 p$. This limit corresponds to the energy-driven phase occurring for $R \gg R_C,$ in which we can safely neglect the Compton cooling term in eq.\ (\ref{eq10}).
By substituting the previous expressions for $M(R), L_w, L_E, $ into eqs. \ref{eq09}-\ref{eq10}, and further assuming that $M_\bullet = M_\sigma$, we find 
\be
R_{s,e} =  \left(\frac{2\eta c \sigma^2}{3}\right)^{1/3} t \equiv v_{s,e} t, 
\label{eq13} 
\ee
i.e. the shell moves at a constant velocity $v_{s,e} = 930 \, \sigma_{200}^{2/3} \kms$ in the isothermal density profile, conserving energy.\footnote{Note that $v_{s,e}\gg \sigma$, thus justifying our assumption that gravity force can be neglected in the energy-driven phase.} 

\subsection{Shell Fragmentation} 

From these asymptotic limits, we can demonstrate that the shell always undergoes a strong acceleration at the transition between the two different driving phases. During the momentum-driven phase, the shell velocity grows $\propto L_E^{1/2}$, i.e with the square root of the black hole mass. The maximum of $v_{s,m}$ is achieved when $M_\bullet\simeq M_\sigma$, i.e just before the transition. In this case, substituting the expression for $M_\sigma$ (eq. \ref{eq08}) into eq.\ (\ref{eq12}), we find $v_{s,m}^{({\rm max})}=\sqrt{2} \sigma$, or
\be
v_{s,m}^{({\rm max})} = 280\, \sigma_{200} \kms \ll v_{s,e} = 930 \, \sigma_{200}^{2/3} \kms.
\label{eq14} 
\ee

The above inequality remains true for any physically reasonable value of $\sigma,$ up to $\approx 7100 \kms$. This fact has an important consequence. It implies that during the acceleration the shell might be disrupted and fragmented by Raleigh-Taylor (RT) instabilities \citep{Rayleigh01111882,1950RSPSA.201..192T}. In fact, the stability condition for the shell is that $d^2R_s/dt^2 <0$ so that the ``effective gravity'' points from the more dilute hot medium into the shell. Stated differently, stability requires $R_s \propto t^\alpha$ with $\alpha <0$. Thus, the constant expansion velocity found in both phases guarantees only marginal stability, and strong RT instabilities are expected to develop during the transition phase. 

The magnitude of the radial acceleration at $R_C$ is 
\be
\vert g_s(R_C) \vert \equiv \ddot R_s \approx \frac{v_{s,e}\Delta v_s }{R_C} \approx \frac{v_{s,e}^2}{R_C}, 
\label{eq15} 
\ee
where we have used the fact that  $\Delta v_s = v_{s,e}-v_{s,m} \approx v_{s,e}$.  The growth rate of the RT instability on spatial scale $\lambda=2\pi/k$ 
\citep{1984PhyD...12...45R, 1991PhFl....3.1312Y, 1998PhRvL..80.3507S, 1999PhPl....6.2009D, 2001JCoPh.169..652G, 2002PNAS...99.2587G,  2004PhFl...16.1668D,Ferrara06, 2008ApJ...686..927S} is given by
\begin{equation} 
\omega_{\rm RT}(\lambda) = 
\left[{{2\pi \vert g_s(R_C)\vert \over \lambda} {(\Delta - 1)\over
      (\Delta + 1)}}\right]^{1/2} \simeq \left[{2\pi \vert g_s(R_C)
    \vert \over \lambda }\right]^{1 \over 2},
\label{eq16}
\end{equation}
the second equality arising from the fact that the density contrast between the shell and the shocked ISM is $\Delta \gg 1$.  Note also that
the instability grows first on the shortest scales and it diverges as $\lambda \rightarrow 0$. At these small scales, the RT is stabilized
by viscosity and/or magnetic fields. Evaluation of  $\omega_{\rm RT}$ at $\lambda = R_C$ finally gives a growth time of the RT instability  
\be
t_{RT} =  \omega_{\rm RT}^{-1} = \frac{1}{\sqrt{2\pi}} \frac{R_C}{v_{s,e}} = 10^5 \,\sigma_{200}^{1/3} \zeta^2_{-1}  M_{\bullet,8}^{1/2} \,\textrm{yr}.
\label{eq17} 
\ee
Hence we see that the RT instability grows on a very short time scale as the shell enters the energy-driven phase. The nonlinear evolution produces a fragmentation of the shell into clumps. The hot medium behind the reverse shock wraps around them; at the same time, the ram pressure on the clumps decreases so they lag with respect to the contact discontinuity. 

\subsection{Shell Fragment Dispersal} 
Once formed, the cold clumps are rapidly ablated and eventually destroyed by the hot material flowing past them.
\citet{Klein94} showed that if hot material moves past a clump of radius $r_c$ at a subsonic velocity, $v_h$ that is much larger than the clump internal sound speed, it will shred it on a ``crushing'' timescale given by
\be
t_{cc} = \frac{\chi_0^{1/2} r_c}{v_h},  
\label{eq18} 
\ee
where $\chi_0$ is the initial density ratio between the clump and the surrounding hot medium.  While this is extended somewhat  in the case in which the exterior medium is moving supersonically with respect to its own soundspeed \citep{Scannapieco15,2016arXiv160701788S}, this is not a big effect in our case in which the hot medium behind the shell is only mildly transonic.

As the clumps originate from the shell,  their sizes must be less than the radius of the shell at the time of break-up,
given by the (maximal) shock cooling length in the momentum-driven phase $r_c \leq R_c  = v_{s,m}^{(\rm max)} t_C$  as given by eq.\ (\ref{eq07}).
The density contrast between the post- and pre-shock material in a radiative shock is equal to the square of the  Mach number  of the shock $ {\cal M}^2 = (v_{s,m}/c_{s,w})^2 $, where we have
neglected the small velocity difference between the shell and the shock and assumed a pre-shock temperature of $10^4$ K to compute the sound speed $c_{s,w}$. On the contrary, the analogous density contrast in the energy-driven phase when the reverse shock becomes adiabatic is only a factor of 4. Hence, $\chi_0^{1/2} \approx  {\cal M}/2$. With these assumptions we get that
\be
t_{cc}  \leq \frac{{\cal M}}{2} \frac{v_{s,m}^{(\max)}}{v_{s,e}} t_C = 3.1 \sigma_{200}^{1/3} \zeta^2_{-1}  M_{\bullet,8}^{1/2} \,\textrm{Myr},   
\label{eq19} 
\ee
which means that the clumps are ablated and mixed with the hot medium on a time that is short with respect to the outflow typical lifetime ($10^{7-8}$ yr),
meaning that they can not contribute to the high velocity, cold and
molecular clumps observed in quasar outflows. On the other hand, the
column depth of the shell at break-up is relatively large $\approx 2
\times 10^{23}$ cm$^{-2},$ meaning that if electron thermal conduction
from the hot medium proceeds at the  (isotropic) Spitzer value, the clumps may be preserved by an evaporative flow significantly longer ($\approx 30 t_{cc}$),  as estimated from the scalings measured in \cite{2016ApJ...822...31B}. Thus, while the broken shell is not likely to contribute to the cold observed gas, the issue of the role of anisotropic conduction during the stage of shell break-up merits further investigation.

\subsection{Shell Fragment Masses Are Small} 
However, regardless of the arguments given on the fragmentation, ablation and final dispersal of the cold clumps, an additional issue prevents their identification with the observed high-velocity molecular components in QSO outflows. This has to do with the limited amount of mass  in the fragments. The forward shock at $R_2$ travels at a speed $\dot R_2 \approx (\gamma+1)/2= (4/3) v_{s,e}$, where $\gamma$ is the ratio of specific heats, taken to be 5/3. The mass outflow rate at the forward shock is then given by 
\be
\dot M_2 = 4\pi \rho_g(R_2) \dot R_2 R_2^2 = \frac{8 f_g \sigma^2}{3G} v_{s,e}.
\label{eq20} 
\ee
Performing the same calculation for the mass outflow rate of the shell, $\dot M_s$ assumed to travel at the maximal velocity $v_{s,m}^{(max)}=\sqrt{2} \sigma$, and taking the ratio of the two we obtain:
\be
\frac{\dot M_2}{\dot M_s} = \frac{4}{3} \frac{v_{s,e}}{v_{s,m}}.
\label{eq21} 
\ee
To obtain the mass ratio between the shocked ISM  and the shell mass we finally need to multiply by the timescale of the two phases. The shell can grow only until $R_C = 0.25 \,\sigma_{200} \zeta^2_{-1}  M_{\bullet,8}^{1/2} \,\textrm{kpc} $, a distance reached after a time $R_C/v_{s,m}$; cooling and condensation of the shocked  ISM  can occur only up to a radius $R_{ff} = 5 \sigma_{200}^{4/3}$ kpc (see next Section) reached after a time $R_{ff}/v_{s,e}$. By substituting these values into eq.\ (\ref{eq21}), we find that the mass ratio is equal to
\be
\frac{M_2}{M_s} = 26.6 \,\sigma_{200}^{1/3} \zeta^2_{-1}  M_{\bullet,8}^{1/2}.
\label{eq22} 
\ee
Thus, most of the outflowing mass is contained in the shock-heated ISM behind the forward shock. 

Finally, the survival of dust grains necessary to catalyze the formation of molecules is much more likely to occur as they pass through the forward shock travelling at a velocity $\dot R_2 \approx 1000 \kms$, rather than across the wind shock (velocity $v_w = 0.1 c$). Thus, in addition to the destruction and mass budget problems of the shell fragments, the lack of dust grains within them  is also a concern, hampering the only efficient channel available to explain the observed molecular content.

\section{Clump Condensation in the Outflow} 

We now explore the alternative scenario in which cold clumps condense out of the shock-heated ambient material behind the \textit{forward} shock at $R_2$ via a thermal instability.
 A parcel of the ISM  at rest engulfed by the forward shock will be heated to a temperature
\be
T_2 =  \frac{3\mu m_p (4/3) v_{s,e}^2 }{16 k_B} = 2.2\times 10^7 \sigma_{200}^{4/3}  \, \textrm{K}. 
\label{eq20fs}
\ee
Assuming that the post-shock medium has a density $4 \rho_a$ and that the cooling function $\Lambda(T) = 3\times 10^{-23} T_7^{-0.7}$ erg cm$^3$ s$^{-1}$ in the range $10^5 \mathrm{K} \simlt T \simlt 10^7 \mathrm{K}$, we find that the cooling time is
\be
t_c = \frac{3}{8}\frac{\mu m_p k T_2}{\rho_a \Lambda(T_2)} = 0.15 \sigma_{200}^{-2} R_{2,\textrm{kpc}}^2 \, \textrm{Myr}.   
\label{eq21fs} 
\ee
This timescale  is shorter than the dynamical time of the shock $t_d \simeq 3 R_2/ 4 v_{s,e} = 0.77 R_{2,\mathrm{kpc}} \sigma_{200}^{2/3}$ Myr only for  $R_{2,\mathrm{kpc}} < 5 \sigma_{200}^{4/3} = R_{ff}$ kpc. Beyond this radius the outer shock becomes adiabatic, thus inverting the standard sequence adiabatic-radiative usually expected for e.g. stellar winds and supernova explosions in quasi-uniform media. This is due to the isothermal stratification of the halo gas which forces the cooling time to increase with radius.
As the two timescales  differ only by a factor of 5, the shock is far from isothermal and it is not clear whether a new shell can form in this intermediate regime. It appears unlikely, though, that the cooling structure of the medium would resemble the thin shell characterizing the momentum-driven phase.

Next, we want to compute the final temperature of the cooling gas, which is regulated by the energy balance between cooling and photoelectric heating by the QSO UV radiation.
We assume that the bolometric luminosity of the QSO is the one derived above, i.e. 
\be
L_E =  \frac{4\pi G c}{\kappa_{es}} M_\sigma = \frac{4f_g c}{G} \sigma^4 = 1.1\times 10^{13} \sigma_{200}^{4} L_\odot. 
\label{eq23} 
\ee 
In order to compute the hydrogen ionizing ($h\nu_L > 13.6$ eV) photon rate, $\dot N_\gamma$, corresponding to the Eddington luminosity eq.\ (\ref{eq23}), we use the bolometric correction, appropriate for the Lyman limit wavelength, 
\be
L_{\rm bol}= {\rm B}_\lambda \lambda L_\lambda,
\ee
with $\lambda_{912} = 912 $A, and  $\BC=4.9$, as given by \citet{Richards06}.
We assume
an unabsorbed continuum blueward of the Lyman limit of the form $L_\lambda = L_{912} (\lambda/\lambda_{912})^{-\alpha}$ with $\alpha=0.5$ as appropriate for a radio-quiet quasar \citep{Telfer02, Mortlock11}, further noting that $\lambda L_\lambda = \nu L_\nu$. Then, 
\be
\dot N_\gamma =  \int_{\nu_L}^\infty \frac{L_\nu}{h\nu} = \frac{\lambda_{912}^2 L_{912}}{h c} \int_0^1 x^{1-\alpha} dx, 
\label{eq25} 
\ee
where $x = \lambda/\lambda_{912}$. By substituting $L_{912} = L_E/\BC \lambda_{912} = 9.9 \times 10^{42} \sigma_{200}^{4} \ergs A^{-1}$ , we obtain the final expression
\be
\dot N_\gamma =   \frac{\lambda_{912} L_E}{\BC h c (2-\alpha)} = 2.7\times 10^{56}  \sigma_{200}^{4} \, {\rm s}^{-1}. 
\label{eq26} 
\ee
This corresponds to a hydrogen photoelectric heating rate  given by
\be
H_\pi = \Gamma_\pi \langle h\nu\rangle_\sigma n_H \approx \frac{\dot N_\gamma \langle h\nu\rangle}{4\pi R^2} \sigma_{H,912}  n_H,
\label{eq28} 
\ee
where $\langle h\nu\rangle_\sigma = (7/5) \,\mathrm{Ryd} = 19$ eV is the
photoionization cross section-weighted mean photon energy for the
adopted QSO spectrum, $\Gamma_\pi= \dot N_\gamma \sigma_{H,912}/4\pi R^2$
is the hydrogen photoionization rate with a cross section at 1 Ryd of
$\sigma_{H,912} =6.3\times 10^{-18}$ cm$^2$. 

For a photoionized gas, one can show that the neutral fraction $x_H =n_H/n$, where $n_H$ ($n$) is the hydrogen (total) number density, is simply written as $x_H
\approx \Xi^{-1} = n\alpha_B(T)/\Gamma_\pi$; $\alpha_B = 6.2\times
10^{-10} T^{-0.845} \mathrm{cm}^3 \mathrm{s}^{-1} \equiv \alpha_0
T^{-0.845} $ is the hydrogen Case B recombination coefficient. Hence,
by substituting, we find that the heating rate goes as 
\be
H_\pi = \langle h\nu\rangle_\sigma \alpha_B(T) n^2 = \frac{\langle h\nu\rangle_\sigma} {t_{\rm rec}} n,
\label{eq28b} 
\ee
which is independent of the photoionization rate  and inversely proportional to the recombination time of the gas $t_{\rm rec} \equiv (n \alpha_B)^{-1}.$

As $\Xi \gg 1$ for $\Gamma_\pi(R=1 \mathrm{kpc})$, $n = n_2$ and any value of $T$, we can assume that hydrogen is fully ionized, $(1-x_H) = x_e \approx 1$. In this case the cooling (in $\ergs$ cm$^{-3}$) is only provided by free-free, for which 
\be
C \approx 1.42\times 10^{-27} \sqrt{T} n^2 \equiv C_0 \sqrt{T} n^2. 
\label{eq29} 
\ee
By imposing $H=C$ we find that the equilibrium temperature is also independent of density, and it is equal to 
\be
T_{eq} =  \left[\left(\frac{\alpha_0} {C_0}\right)\langle h\nu\rangle_\sigma\right]^{1/1.345} = 1.9\times 10^5 {\rm K}.
\label{eq30} 
\ee
Thus, the unattenuated QSO radiation field would prevent the shocked ISM from cooling back to temperatures where molecules could form. This conclusion however must be interpreted as an upper limit to the temperature because while hydrogen is mostly ionized as we have seen, heavier elements are likely to be only partially ionized and can continue to contribute to the cooling. 

The above argument shows that indeed there is a ``window of opportunity'' for the gas heated by the forward shock to radiatively cool via two-body interactions in a region extending from $R_C$ to about 5 kpc. Whether this cooling leads to the formation of clumps via a thermal instability can only be assessed by numerical simulations. If these clumps indeed form, they will be born with the same velocity of the hot gas and therefore they would not suffer from ablation as those resulting from the momentum-driven shell fragmentation discussed above. Therefore they could persist and reach distances in excess of 10 kpc, in agreement with recent experimental findings \citep{Feruglio10, Cicone15}.

\section{Simulations}

\subsection{Initial Conditions}

To study thermal instabilities occurring in the shock-heated ISM gas behind the forward shock in further detail,  we carried out a set of four simulations using FLASH (version 4.2), a multidimensional hydrodynamics code \citep{2000ApJS..131..273F}.  For these we focused on the case in which $\sigma_{200}=1$ and $M_{\bullet,8} = 1,$ and for simplicity we took the velocity of the forward shock to be time-independent $\dot R_2 = (4/3) v_{s,e} = 1240 \kms,$ in which case the gas behind the shock is heated to a temperature $T_2=2.2\times 10^7 \textrm{K}.$  We started the computation well into the energy-driven phase in which $R_2 > R_C = 0.25 \,\sigma_{200} \zeta^2_{-1}  M_{\bullet,8}^{1/2} \,\textrm{kpc} $, and therefore chose the initial radius as $R_0= 1$ kpc. This maximized the formation chances of clumps as $t_c$ rapidly increases towards larger distances from the central black hole due to the declining density.  The density of the gas at this radius is $\rho_2 = 4\rho_a(R=1 {\rm kpc}) = 6.4\times 10^{-23}$ g $\cc$, or $n_2 = \rho_2/\mu m_p = 60 \cc$.

The simulations used periodic boundary conditions and captured the evolution of a representative region 1 kpc on a side, made up of  $512^3$ cells with a  fixed resolution of $\Delta x = 1.95$ pc. As discussed below, this allowed us to resolve the Field length with $\approx 10$ cells. The ambient ISM gas was assumed to have solar metallicity ($Z=Z_\odot$), and a Galactic dust-to-gas ratio,  $\rho_d/\rho_a = {\cal D} = {\cal D}_\odot = 1/100$. 

To mimic the small-scale density fluctuations present in the gas, we imposed random isothermal fluctuations in log density with a white noise power spectrum, down to a scale of $2 \Delta x = 3.9$pc, such that for each $2 \Delta x = 3.9$pc cubic region, 
\be
n({\bf x}) = n_0 \frac{2 a}{e^A - e^{-A}} e^{(A r)},
\ee
where $r$ is a random number between $-1$ and $1$, $n_0 =  60\, {\rm cm}^{-3},$ and the amplitude of the fluctuations is set by $A$.  Note that in the limit of small $A$ these are equivalent to linear fluctuations with an rms value of $n_0 a (2/3)^{1/2}$, but this formulation allows for rms fluctuations of arbitrary amplitude. The perturbations are the seed for the initial stages of the evolution of the medium, which is determined by  atomic cooling and heating processes, conduction, and dust cooling and destruction.

In this study, we used the unsplit flux solver with a predictor-corrector type formulation based on the method presented in \cite{2013JCoPh.243..269L}, enabling an adaptively varying-order reconstruction scheme  that reduces its order to first-order depending on monotonicity constraints. We also made use of the shock detect flag which lowered the prefactor in the Courant-Friedrichs-Lewy timestep condition from its default value of 0.4 to 0.25 in the presence of strong shocks.

\subsection{Conduction}

Two of the simulations described below included 
electron thermal condition,  a strongly temperature-dependent process that goes as $ d E/dt =  \nabla {\bf q} $, where the energy flux is
\be
{\bf q} = {\rm min} \, \begin{cases}
\kappa_e(T_e) \nabla T_e\\
0.34 n_e k_{\rm B} T_e  c_{\rm s,e},
\end{cases} 
\label{eq:thermalcond}
\ee
where $\kappa_e(T_e) = 5.6 \times 10^{-7} T_e^{5/2}$ erg s$^{-1}$ K$^{-1}$ cm$^{-1}$  and  $c_{\rm s,e} = (k_{\rm B} T_e/m_{\rm e})^{1/2}$ is the isothermal sound speed of the electrons with a temperature $T_e$ \citep{1977ApJ...211..135C}.
This leads to a flux limited-diffusion equation which we solved using the general implicit diffusion solver in FLASH, assuming that the electrons and ions had the same temperature.   The solver uses the HYPRE\footnote{https://computation.llnl.gov/casc/hypre/software.html} linear algebra library to solve the discretized diffusion equation, modifying local the diffusion coefficient to vary smoothly up to $0.34 n_e k_{\rm B} T  c_{\rm s,e}$, where we have used the Larsen flux limiter \citep{2000JQSRT..65..769M}. 

In the presence of conduction,  thermally unstable clumps of cool gas will collapse to the Field length, i.e. the length-scale below which thermal conduction suppresses local thermal instability \citep{1965ApJ...142..531F}. This is given by
\be
\lambda_F =  \sqrt{\frac{\kappa_e(T_2) T_2}{n_2^2 \Lambda(T_2)}} = 25.7 \,\textrm{pc}, 
\label{eq:condflux} 
\ee
where we have taken $T_2 = 2.2 \times 10^7$K and $n_2$ = 60 as above.  Thus our simulations are able to resolve this scale with $\approx$ 10 cells. 

\subsection{Atomic / Ionic Cooling and Heating Process}

As discussed above, once it is in the energy-driven phase, the black hole mass  is likely to stabilize to the value that appropriately puts it onto the $M_{\bullet}-\sigma$ relation. Therefore during the simulated evolution, we kept the bolometric luminosity of the QSO at the corresponding Eddington rate as given by eq.\ (\ref{eq26}). As described in \S 2.4, atomic and ionic cooling in the post-shock gas is complicated by the fact that the material is strongly illuminated by the QSO radiation field, which plays a key role in the thermodynamic evolution of the gas by: (i)  changing the ionization fractions of the various elements and thus greatly changing the rate at which they radiate energy through line emission; (ii) providing a heating source through the kinetic energy imparted to  photoionized electrons, and (iii) providing an additional heating source through  Compton scattering with free electrons.    

In order to deal with the first two of these effects, we used the recent computations of atomic and ionic cooling/heating functions of photoionized gas performed by \citet{Gnedin12} \citep[see also][]{Sazonov05}.  This allowed us to account for both the heating from photoionization as well as changes in radiative cooling that occur as species are shifted to higher ionization states than they would be in the absence of a background.

\begin{figure}
\center{\includegraphics[width=80mm]{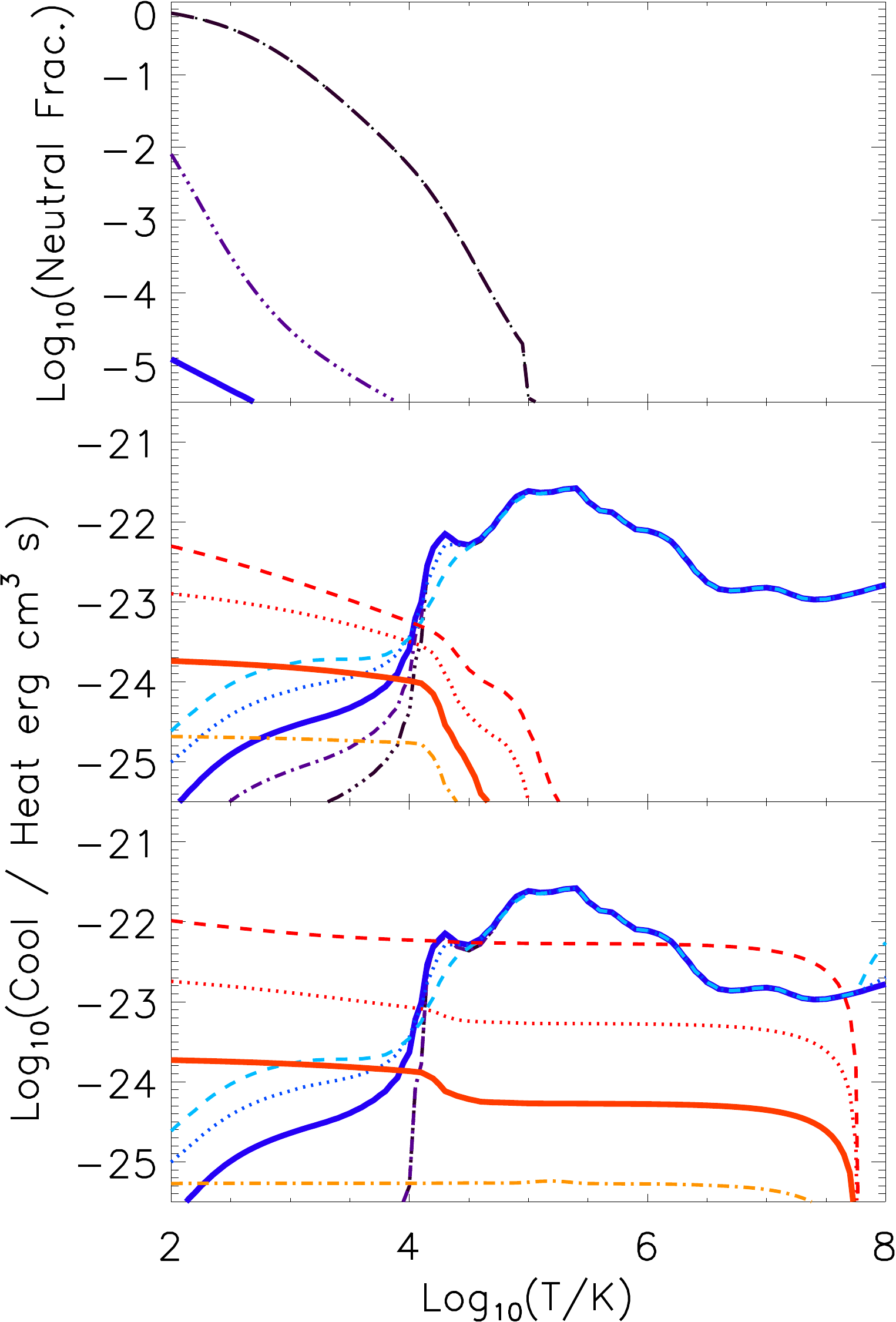}}
\caption{ {\em Top:} Neutral fraction  as given by eq.\ (\ref{eq50}). {\em Middle:} Radiative cooling rates (blue lines)  vs. photoheating rate (red lines) for optically thin material, normalized by the mean baryonic density squared.    {\em Bottom:}  Radiative cooling rates and heating ratios including the attenuation of the ionizing flux according to eq.\ (\ref{eq43}) as well as Compton heating/cooling as given by eq.\ (\ref{eq34a}). In all panels, the solid lines are for material with a mean density of $\rho/\mu m_p$ = 60 cm$^{-3}$, while the dashed, dotted, dot-dashed, and triple-dot-dashed lines correspond to material with $\rho/\mu m_p$ = 0.6, 6, 600, and 6000 cm$^{-3}$, respectively.} 
\label{Fig2}
\end{figure}

In Figure \ref{Fig2}, we show the cooling and heating rate normalized by the mean baryonic number density, $n_B^2,$ computed for a range of particle densities  from 100 times to 1/100 of the mean density in our simulations.  At the higher densities, $\geq$ 60 cm$^{-3}$, recombination times are  short  and the cooling rate is similar to the case without a background.  In this case photoheating is limited by the ionization rate rather than the recombination time, such that the normalized heating rate is negligible.  

 At densities $\leq$ 6 cm$^{-3}$, on the other hand, atomic/ionic cooling and heating are both significantly affected by the presence of the background.    In the case of cooling, the lowest temperature peak of the cooling function is significantly lessened, due primarily to hydrogen remaining almost completely ionized.   The hydrogen ionization rate has been boosted to the point where it is comparable to the recombination rate, leading to a heating rate close to the limit described by eq.\ (\ref{eq28}). Together these effects shift the point at which cooling and heating are balanced from $\approx 8000$ K for gas with  $\rho/\mu m_p$ = 60 cm$^{-3}$ to  $\approx  1.5 \times 10^4$K for gas with  $\rho/\mu m_p$ = 0.6 cm$^{-3}.$
 
However, these curves are not sufficient to fully treat the impact of the QSO radiation field, as they  make two significant approximations, which can have important effects on the thermodynamic evolution of the gas.  First, they do not consider self-shielding, which can happen when clumps that form within the gas develop a high-enough optical depth to maintain a significant neutral fraction at their centers, and second they do not include heading by Compton scattering.  To estimate self-shielding we calculated the neutral fraction of hydrogen according to eq.\ (\ref{eq50}), yielding results plotted in the top panels of Fig.\ \ref{Fig2}. Assuming $\approx 10$pc clumps, corresponding to the Field length at the mean density, as described above, we are then able to associate an optical depth with each density in our simulation.  Finally, making use of the fits to detailed radiative transfer simulation performed by \citet{Rahmati13}, we used these values to examine the mean attenuation of the background as a function of gas as described in Appendix B.

To estimate the contribution from Compton heating arising from the QSO radiation field at temperatures below the Compton temperature, $T < T_C$, we boosted the overall heating rate of the gas by a factor of
\be
H_C= \frac{\sigma_T F}{m_e c^2}  (\langle h\nu\rangle -4 k_B T) n_e =\Gamma_C \Delta\epsilon \,n_e. 
\label{eq34a} 
\ee
Here $T_C \equiv \langle h\nu \rangle/4 k_B$,  where $\langle h\nu \rangle = L^{-1} \int_a^b h\nu L_\nu d\nu$. If, as in eq.\ (\ref{eq25}) we assume that the spectral shape of the QSO is $\L_\nu \propto \nu^{\alpha-2}$ with $\alpha=0.5$, for $(a, b) =(13.6 \, \mathrm{eV}, 1 \, \mathrm{GeV})$ we find  $\langle h\nu \rangle = 20$ keV or $T_C=5.8\times 10^7$K.  Note, however, that while the precise value of $T_C$ is sensitive to the high energy integration limit and spectral slope, it will primarily affect the temperature of the hot intercloud medium, while we are interested in the cold phase (clumps), and thus this uncertainty only weakly impacts our results. Including these two effects shifts the cooling and heating curves from their \citet{Gnedin12} values to those shown in the lower panels of Figure \ref{Fig2}.  By comparing the cooling curves in this panel to those in the middle panel, we can see that, as expected, self-shielding effects are only important at the very highest densities, for which large neutral fractions lead to substantial optical depths, moving the 600 cm$^3$ and 6000 cm$^{-3}$ curves on top of each other.

On the other hand, as the energy input from Compton scattering  is  $\propto n_e$ rather than density squared, it  has the largest impact at the lowest densities, becoming so large in the 0.6 cm$^{-3}$ case as to exceed the cooling rate for all gas with $T \geq 10^6$K.   At these low densities, a hot inter-clump medium can be sustained indefinitely, as explored in further detail below.   Moving to higher densities, Compton effects are still important at 6 cm$^{-3}$, but have little effect at densities at or above 60 cm$^{-3}$, falling well below the atomic/ionic cooling rates at high temperatures, and well below the photoheating rates at low temperatures.  

\subsection{Dust Cooling and Destruction}

A final complication to the evolution of the outflowing material is the presence of dust, which is able to also provide additional mechanisms  for cooling as well as heating, although the latter is negligible compared to the photoelectric heating by the QSO. When the thermal energy of electrons ejected from the grain surface is less than the energy of electrons recombining on positively charged grains, as is the case when a strong UV field is present,  there is a net cooling of the gas. This cooling rate depends on the flux in the UV (``Habing'') band 6-13.6 eV, $G_0=F/F_H$, where the source flux is normalized to the Habing flux $= 1.6\times 10^{-3} \ergs$ cm$^{-2}$. For our QSO we take $F=L_\lambda\Delta \lambda/4 \pi R^2$, with $\Delta \lambda$ the wavelength width of the Habing band (1163 A), which corresponds to $G_0 = 45180$ in our fiducial model. Then, the cooling rate due to recombinations on dust for a solar metallicity gas with Milky Way dust-to-gas ratio, $ {\cal D}$, is
\be
C_d^{\rm rec} \approx 4.65\times 10^{-30} \, \ergs {\rm cm}^{-3} \, \left(\frac{{\cal D}}{{\cal D}_\odot}\right) T^{0.94} \left(\frac{G_0 \sqrt{T} }{n}\right)^{\delta} n n_e, 
\label{eq31} 
\ee
with $\delta=0.74/T^{0.068}$. Taking $T^{0.068} \approx 2$ in the relevant range we can estimate the equilibrium temperature between dust cooling and photoelectric heating as 
\be
T_{eq} =  \left[\left(\frac{\alpha_0} {C_1}\right)\langle h\nu\rangle_\sigma\right]^{1/(1.785+\delta/2)} = 1.2\times 10^4  {\rm K}
\label{eq32} 
\ee
with $C_1=4.65\times 10^{-30} (G_0/n_2)^\delta$. This result clearly shows that dust cooling, if grains survive the passage through the forward shock, dominates over free-free and might be competitive with radiation losses from heavy elements, if the latter are substantially ionized.

In addition to recombination cooling, dust grains can also radiatively cool the gas. Thermal energy from the gas is transferred to the grains by ion-grain and electron-grain collisions, and subsequently radiated in the infrared by the dust. The cooling rate due to this mechanism has been computed by the classical works by
\citet{1981ApJ...245..880D} and \citet{1981ApJ...248..138D} and can be written as $C_d^{\rm rad} = n_g H_{coll}$
(in $\ergs$ cm$^{-3}$) where $n_g ={\cal D}_0 n_2 (\mu m_p/m_{g,0}) $ is the number density of dust particles (assumed to be spherical with radius $a$) in the gas, and 
\be
m_{g} = 4\pi a^3\delta_g/3,
\label{eq:grainmass}
\ee
is the grain mass. The heating rate deposited by particles in the grain can be conveniently parametrized (Montier \& Giard 2004) as
\ba
H_{coll} (a, T, n_e) = \begin{cases} 
5.38\times 10^{-10} n_e a^2 T^{3/2} &  x^* >4.5\\
1.47\times 10^{-3} n_e a^{2.41} T^{0.88} & x^* > 1.5\\
6.48\times 10^{6} n_e a^3 & x^* \le 1.5,
\end{cases} 
\label{eq32a} 
\ea
where $x^* = 1.26\times 10^{11} a^{2/3}/T$. 

We include dust as a separate fluid in the simulation, following its kinematical and mass evolution  in the shock-heated ISM using a passive tracer field  that travels with the flow, neglecting dynamical effects such as radiation pressure and gas drag. In each cell we initially set a value for ${\cal D} = {\cal D}_0 $ and assume a single-grain population, whose pre-shock radius is taken to the average one in a \cite{1977ApJ...217..425M} size distribution, $\langle a \rangle = 0.1 \mu.$

The subsequent grain erosion is a two-step process. First, the grains  traverse the forward shock wave, traveling at a velocity $3/4 \dot R_2$ relative to the post-shock gas. The drag forces due to collisions with the gas particles will slow them down, finally setting them at rest, but also eroding them by sputtering. Detailed calculations of the destruction process are available in the literature, \citep[e.g.][]{McKee87, Jones94, Dwek96, Bocchio14, Slavin15}.            
\begin{figure}[t]
\centerline{\includegraphics[width=90mm]{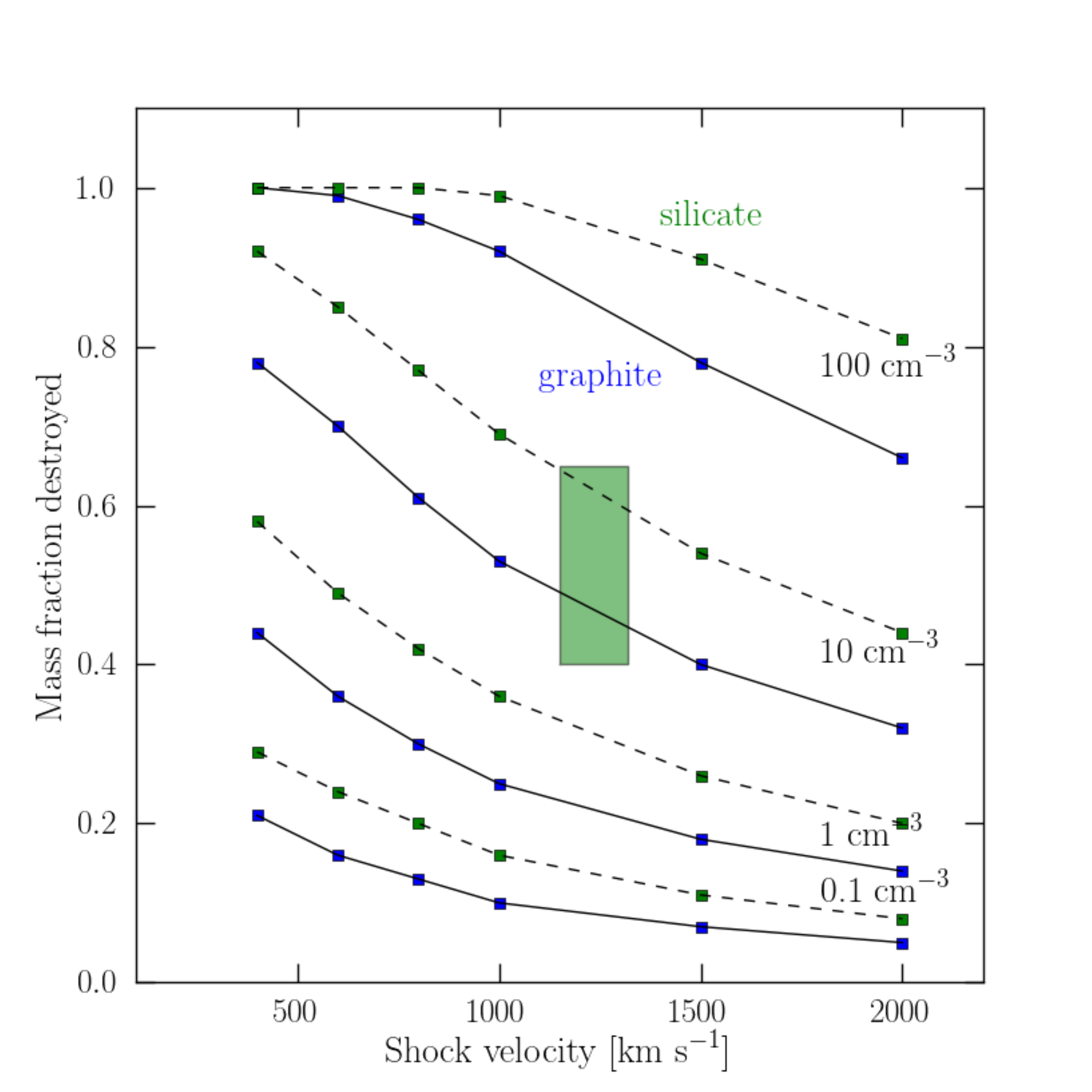}}
\caption{Mass fraction of dust destroyed as a function of shock velocity and ambient density. Data from \citet{Dwek96}. Indicated (green rectangle) is also the parameter region for our simulated case in which the shock velocity is 1240 $\kms$ and the ambient gas density is $15 \cc$.       
} 
\label{Fig03}
\end{figure}

In Figure \ref{Fig03}, we show the results from the \citet{Dwek96} dust destruction model, relative to a fast non-radiative shock. We choose this set of results as the outflow shock is only mildly radiative and because the model includes shock velocities up to $2000 \kms$ of interest here. \citet{Dwek96} predicts that for $\dot R_2 = 1240 \kms$ and an ambient density of $n_a=15 \cc$, the mass fraction of graphite (silicate) grains\footnote{Most of the mass returning to the gas phase is provided by grains $\ll 0.1 \mu$m, thus our choice of the grain size is broadly physically consistent.} destroyed is $1-\gamma_d^g= 0.48$ ($1-\gamma_d^s = 0.62$). As we assume a 1:1 mixture of these two materials, the amount of dust surviving the passage through the QSO outflow is $\langle \gamma_d \rangle =0.45$. The \citet{Dwek96} model follows the grain erosion only for the short time necessary to the grains to travel about 90\% of the shock swept-up mass. For the shock velocity and ambient density of the outflow, this time is about 340 yr. Thus we can consider this first destruction step as ``instantaneous''. Hence we initialize the dust-to-gas ratio to a value ${\cal D}_0 = \langle \gamma_d \rangle {\cal D}_\odot$ and the dust radius to  $a_0 = 0.1 \mu \langle \gamma_d \rangle^{1/3}.$

As the grains are subsequently mixed with the hot post-shock gas, they suffer a more gradual erosion due to thermal sputtering. At this time, the rate at which the grain radius decreases is described by a simple fit to the \citet{Dwek96} results, which are very close to those from \citet{Draine79} and \citet{Tsai95}:
\be
\frac{da} {dt} = - A n T_6^{-1/4} e^{-BT_6^{-1/2}},
\label{eq33} 
\ee
where $n$ and $T$ are the cell by cell ISM density and temperature (in units of $10^6K$), and we have adopted material-averaged values for the constants $(A, B)=(1.2\times 10^{-5} \mu$m yr$^{-1}, 3.85)$. The sputtering rate drops rapidly below $10^6$ K and therefore grain survival critically depends on the growth rate of the thermal instability producing the cold clumps in which grains are finally guarded. 

Using eq.\ (\ref{eq33}), the rate of decrease of dust mass of a single grain can be derived as
\be
\frac{dm_g} {dt} = 4\pi a^2 \delta_g\frac{da} {dt},
\label{eq34} 
\ee
where the grain density $\delta_g = 3$ g $\cc$. 
We implement the grain destruction physics in the simulation by assuming that the pre-shock size of the grains is $\langle a \rangle = 0.1 \,\mu$m,   everywhere and that the initial grain number density is constant throughout the simulation. This allows us to track the evolution of the dust mass fraction as
\be
\partial_t \rho D + \nabla\cdot\left(D \rho {\bf v}\right) = \frac{\rho D_0}{m_0} \frac{dm_g} {dt} ,
\ee
where $D_0$ is the initial dust mass fraction and $m_0 = (4 \pi/3) \delta_g a_0^3.$

\begin{figure*}
\center{\includegraphics[width=145mm]{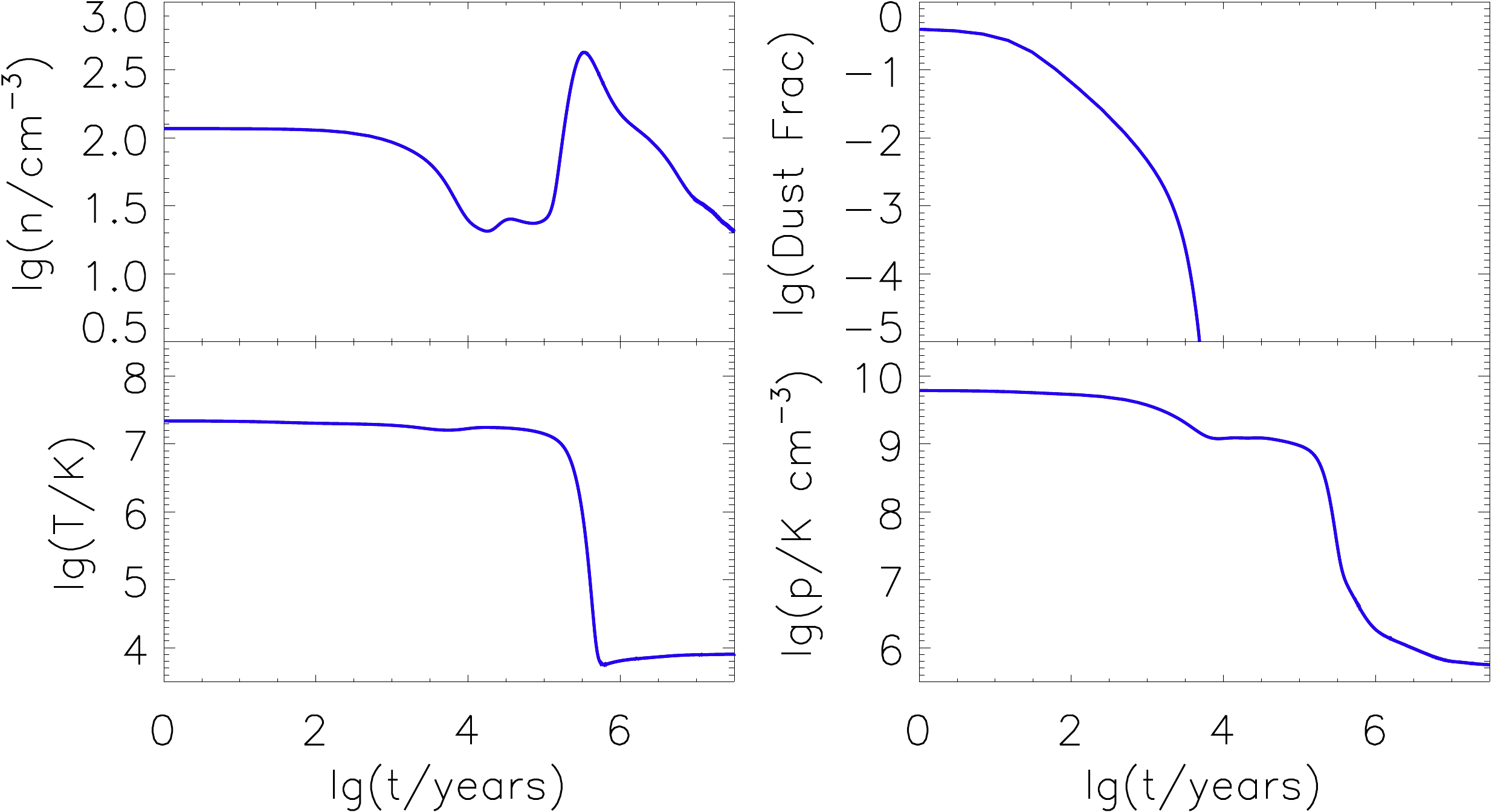}}
\caption{Evolution of thermodynamic quantities in our fiducial model as a function of log time in our fiducial simulation.  {\em Top Left:} RMS density fluctuations  {\em Top Right:} Dust density in units of the initial (pre-shock) density.   {\em Bottom Left:} Mass-averaged temperature. {\em Bottom Right:}  Mass-averaged pressure.}
\label{fig:run1evol}
\end{figure*}

\begin{figure*}
\center{\includegraphics[width=160mm]{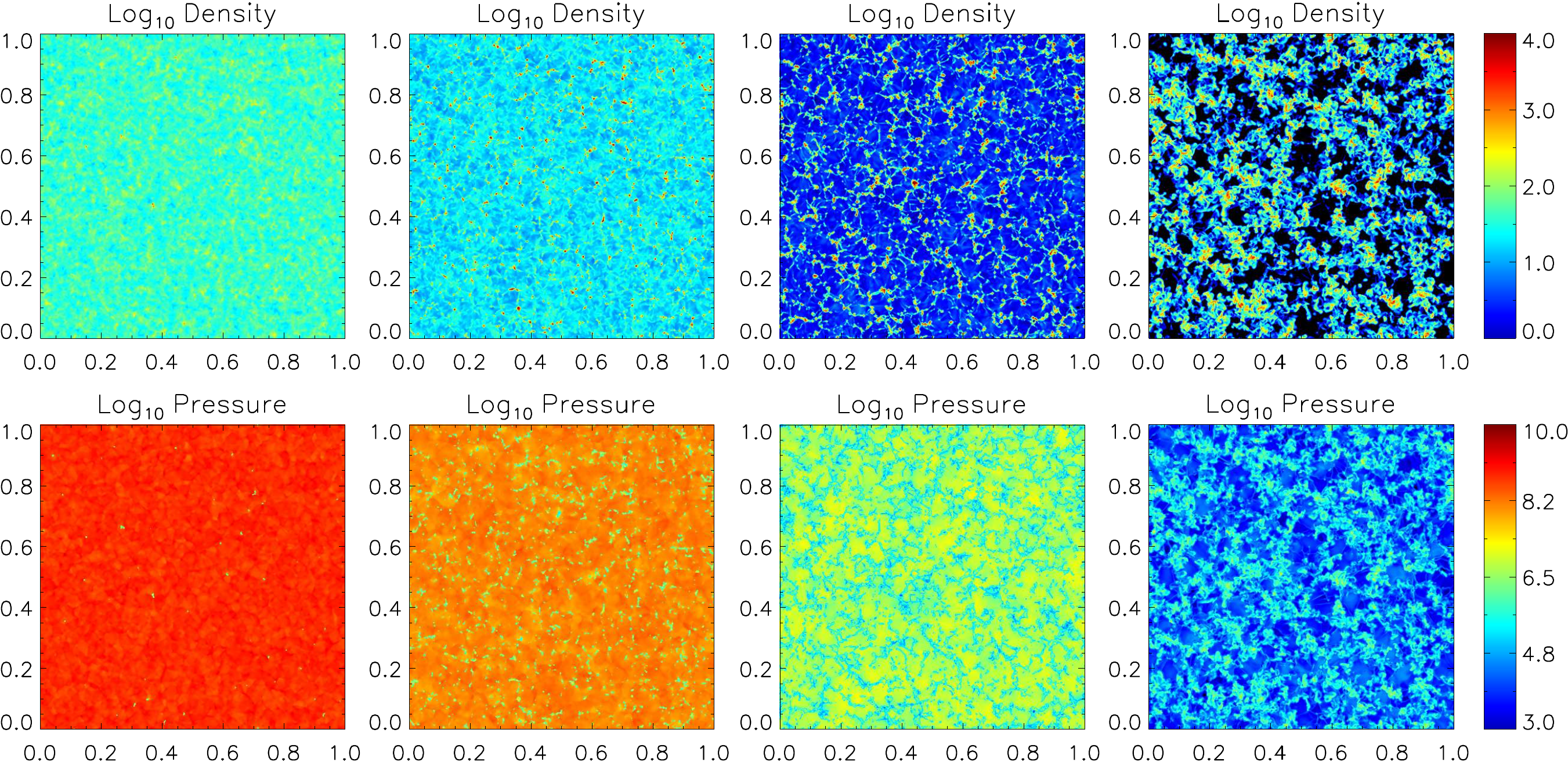}}
\caption{Snapshots of number density in units of cm$^{-3}$ ({\em Top Row}) and pressure  in units of K cm$^{-3}$  ({\em Bottom Row}) on a slice from our fiducial simulations, at characteristic times of 0.3 Myr ({\em First Column}), 1.0 Myr ({\em Second Column}), 3.0 Myr ({\em Third Column}), and 10 Myr ({\em Fourth Column}).  All distances are in kpc.}
\label{fig:run1slice}
\end{figure*}

\section{Results}

Having implemented the models for cooling, heating, conduction, and dust physics described above we then carried out a set of four simulations designed to better elucidate the physics of formation of cold clumps in QSO systems.

\subsection{Fiducial Case Without Electron Thermal Conduction}

In our first simulation, we set the QSO background to its fiducial value, but did not include conductive effects.  In order the maximize the chance for the system to form structures, we assumed large  (small scale) initial density fluctuations with an RMS amplitude of 120 cm$^{-3}.$   The evolution of the  RMS density fluctuations, dust density, mass-averaged temperature, and pressure are shown in Fig.\ \ref{fig:run1evol}, and snapshots of slices of number density and pressure at characteristic times are shown in Fig.\ \ref{fig:run1slice}.

At $2.2 \times 10^7$K, the initial sound speed in the system is $710 \kms$, leading to rapid smoothing of the initial perturbations within $\approx 10^4$ years.  During this short time, rapid dust cooling is able to reduce the gas temperature by about 20\%.  However, sputtering of the dust from the hot medium leads to dust destruction on a similar timescale.   The removal of this coolant stalls gas cooling for $\approx 3 \times 10^5$ years until the significantly longer times at which line cooling and bremsstrahlung processes are able to operate.  During this period, the amplitude of density fluctuations drops to $\approx 25$\%, down by a large factor from the initial value.

Once cooling begins to operate efficiently, however, the temperature of the gas  drops rapidly, beginning with the densest regions.   As underdense regions are over-pressurized with respect to cooler clumps, a pressure wave compresses the dense regions further, leading to a rapid increase of density fluctuations to RMS amplitudes greater than 300 cm$^{-3}$.   

Finally, as cooling becomes more widespread, the gas throughout the simulation drops to the roughly constant $\approx 8000$ K temperature floor set by the balance between photo-heating and atomic cooling.  At this point the pressure gradients switch, such that once again the densest regions of the simulations are the ones with the highest pressure.   At this stage, the clumps begin to dissolve, leading to RMS density fluctuations that decrease rapidly and continuously, falling to $\approx 20$ cm$^{-3}$ at a time of 30 Myr.  Thus, clumps represent only a transient phase during the outflow evolution.   

Perhaps surprisingly, this steady tendency of the gas cool to a single phase occurs even in the presence of the Compton heating provided by the QSO radiation field. This is because, even though a two-phase medium is possible at $\approx 10^6$ K cm$^{-3}$ (as shown in Appendix A),    by the time this pressure is reached at the end of the simulation, there are simply no regions left with low enough densities to move towards the hot stable phase depicted in Fig. \ref{Fig13}.  Note this evolution towards a uniform single-phase medium occurs even with our choice of a vary large initial RMS amplitude of the fluctuations. This suggests that it is extremely difficult for a two-phase medium to develop in a QSO outflow such as the one described here.

\subsection{Electron Thermal Conduction}

\begin{figure*}
\center{\includegraphics[width=145mm]{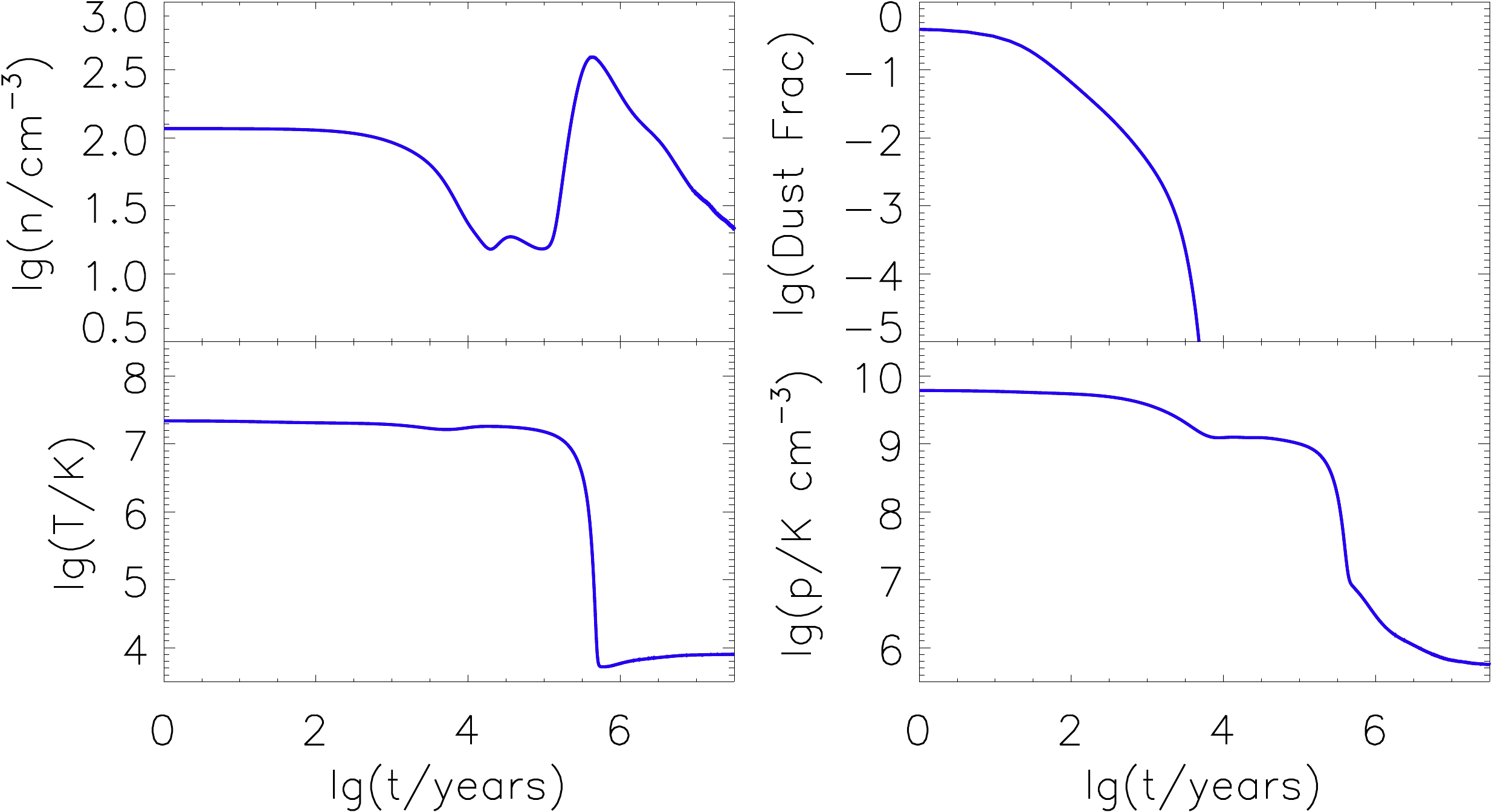}}
\caption{Evolution of thermodynamic quantities in a model with electron thermal conduction included.  Panels are as in Figure \ref{fig:run1evol}.}
\label{fig:run3evol}
\end{figure*}

\begin{figure*}
\center{\includegraphics[width=160mm]{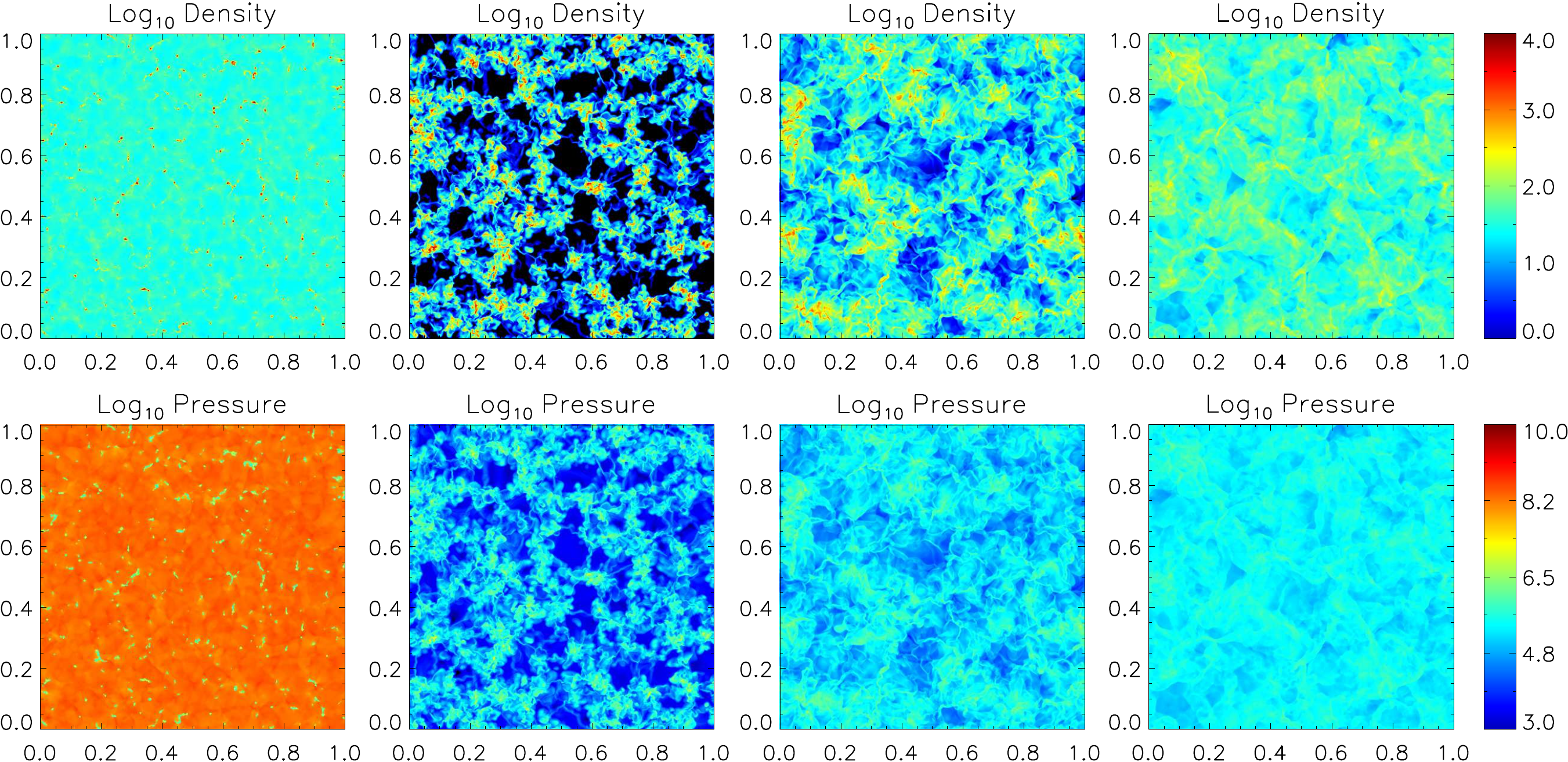}}
\caption{Snapshots of number density in units of cm$^{-3}$ ({\em Top Row}) and pressure  in units of K cm$^{-3}$  ({\em Bottom Row}) on a slice from our simulation with electron thermal conduction, at characteristic times of  of 0.3 Myr ({\em First Column}), 1.0 Myr ({\em Second Column}), 3.0 Myr ({\em Third Column}), and 10 Myr ({\em Fourth Column}).}
\label{fig:run3slice}
\end{figure*}

In a second simulation, we repeated our fiducial case, but now also including electron thermal conduction. This leads to the time evolution shown in Fig.\ \ref{fig:run3evol}, and the  corresponding slices of number density and pressure shown in Fig.\ \ref{fig:run3slice}.   Like the run without conduction, there is an initial phase of rapid cooling which is terminated by the destruction of the dust, and like the run without conduction, this occurs simultaneously with a rapid decrease in the magnitude of the density fluctuations.  In this case however, the presence of conduction leads to much more uniform conditions in the rarified medium, punctuated by dense clumps, whose larger internal energy per unit volume causes them to be less affected by conduction than their surroundings.

As in the case without conduction, these densest regions cool first, and the pressure of the exterior medium compresses them further.   By $\approx 1$ Myr this leads to very large density contrasts arranged into a network of clumps and filaments, whose appearance is not unlike those observed in cosmological simulations, although in this case they are shaped by pressure gradients rather than by gravity.    This arrangement is short-lived however, as the gas cools to $\approx 8000$ K throughout the simulation, pressure gradients reverse, and as in the fiducial case, density contrasts are rapidly washed away as the thermodynamic structures converges towards a single, cold phase.     Again, this motion towards a uniform single-phase medium occurs even in the presence of significant Compton heating and extremely large initial density contrasts, raising the question of just how extreme conditions must become to sustain a two-phase medium in a QSO outflow.

\subsection{Enhanced Compton Heating}

To address this question, we carried out two additional runs, each with a Compton background  whose flux was artificially boosted by a factor of 100, but with all other simulations parameters left unchanged.  In this case, the minimum pressure at which two phases can exist simultaneously is raised by two orders of magnitude to $\approx 3 \times 10^8$ K cm$^{-3}.$  

\begin{figure*}
\center{\includegraphics[width=140mm]{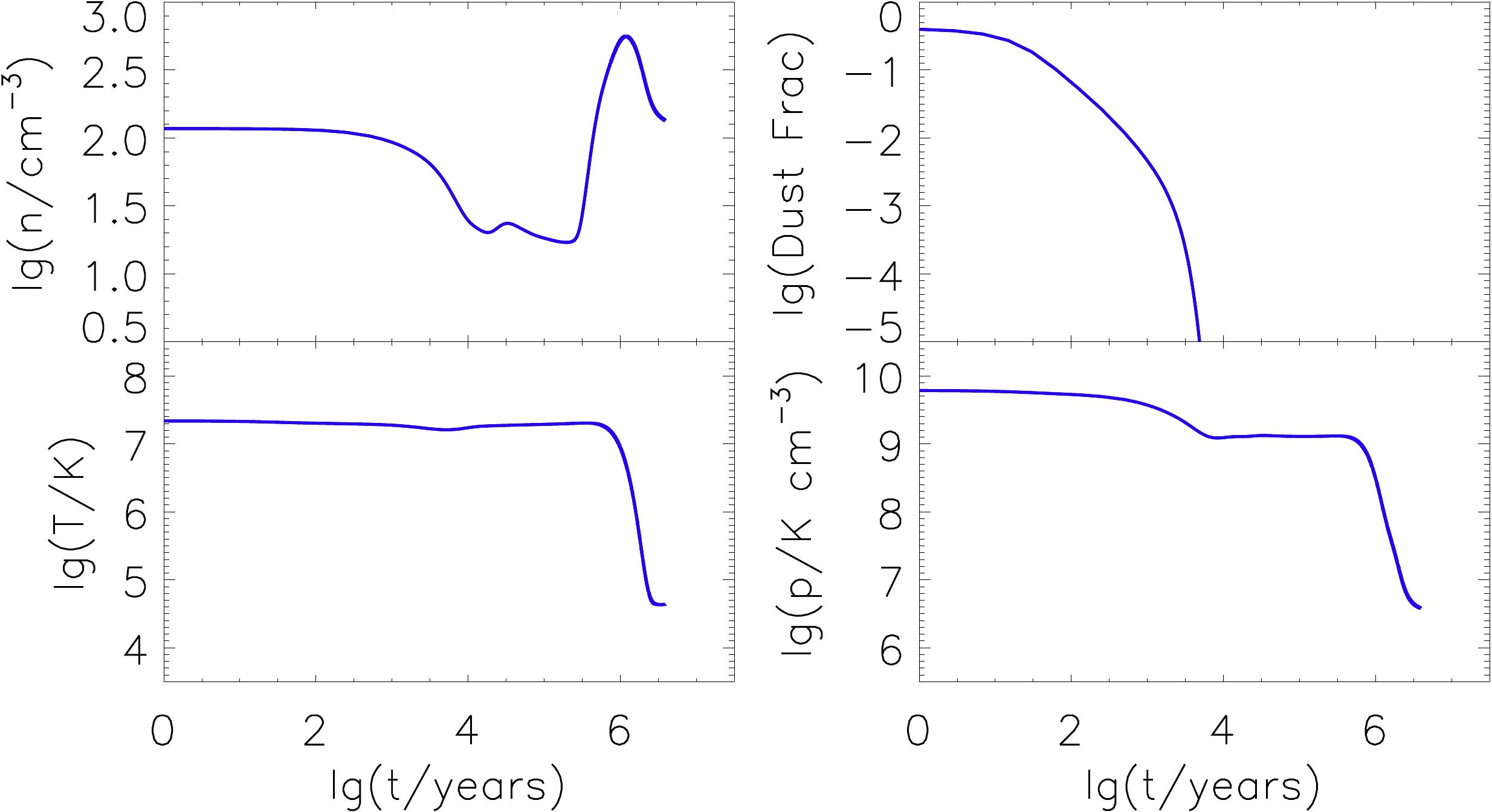}}
\caption{Evolution of material in a model without electron thermal conduction and the Compton background raised by a factor of 100.   Panels are as in Figure \ref{fig:run1evol}, except for the panel illustrating temperature evolution, where now both the density weighted temperature (solid blue line) and the volume weighted temperature (dashed red line) are shown.}
\label{fig:run2evol}
\end{figure*}

\begin{figure*}
\center{\includegraphics[width=160mm]{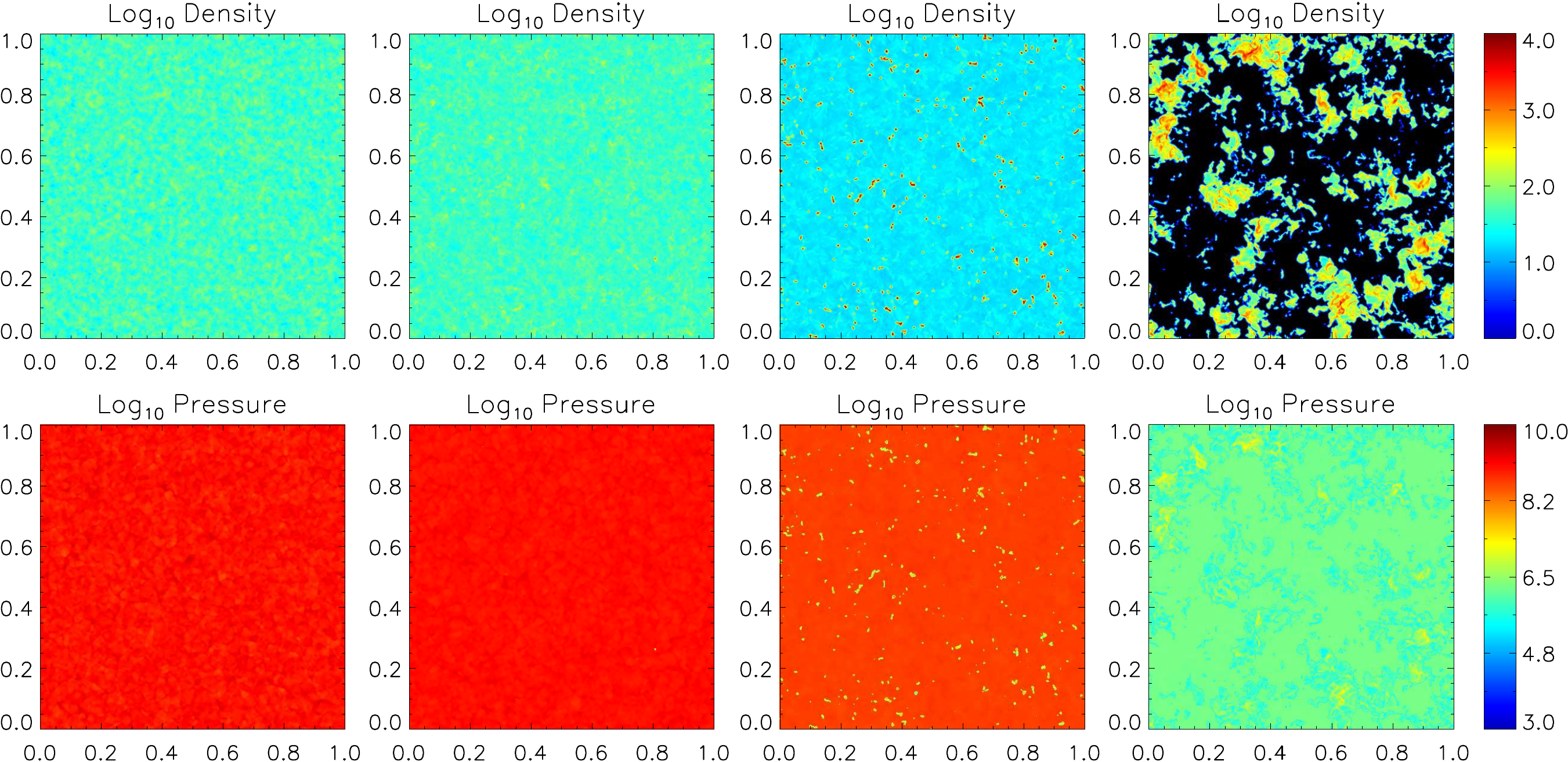}}
\caption{Snapshots of number density in units of cm$^{-3}$ ({\em Top Row}) and pressure  in units of K cm$^{-3}$  ({\em Bottom Row}) on a slice from a simulation without electron thermal conduction and the Compton background raised by a factor of 100. Panels are at characteristic times of 0.1 Myr ({\em First Column}), 0.3 Myr ({\em Second Column}), 1.0 Myr ({\em Third Column}), and 3. Myr ({\em Fourth Column}). Note that these times are earlier than the snapshots in Figures \ref{fig:run1slice} and \ref{fig:run3slice}.}
\label{fig:run2slice}
\end{figure*}

In the  first of these runs we neglected electron thermal conduction, leading to the results shown in Figures \ref{fig:run2evol}  and \ref{fig:run2slice}.    The early stages of this simulation remain the same as in the first two cases, with dust cooling rapidly reducing  the gas temperature by about 20\% during the first $10^4$ years, until rapid sputtering destroys the dust.  Also like the cases above, this phase is accompanied by a strong drop in density contrasts, as pressure differences act to erase the largest density peaks, followed  by a rapid increase in density contrasts, as rapid cooling in these clump drops their pressure well below the pressure in the surrounding medium.

Unlike previous runs, however, the temperature of the rarefied gas during this phase remains high, as Compton heating is able to exceed cooling in all gas with densities below $\approx 3 \times 10^8$ K cm$^{-3} $/($2 \times 10^7$ K) =15 cm$^{-3}.$  This leads to a distribution of cold, low pressure clumps with densities $\approx 1000$ cm$^{-3},$ which are embedded in a largely-uniform rarified medium.  As the gas within these clumps cools to the temperature floor of $\approx 8000$ K, they accrete even more material, until the exterior medium becomes extremely rarified.  At this point the mass-weighted average temperature deviates strongly from the volume weighted temperature. This is because  the cold phase contains almost all of the mass, but the hot phase fills most of the volume while containing only a small fraction of the mass.  In this case, a stable two-phase medium configuration, indeed, is achieved by the system. 

\begin{figure*}
\center{\includegraphics[width=145mm]{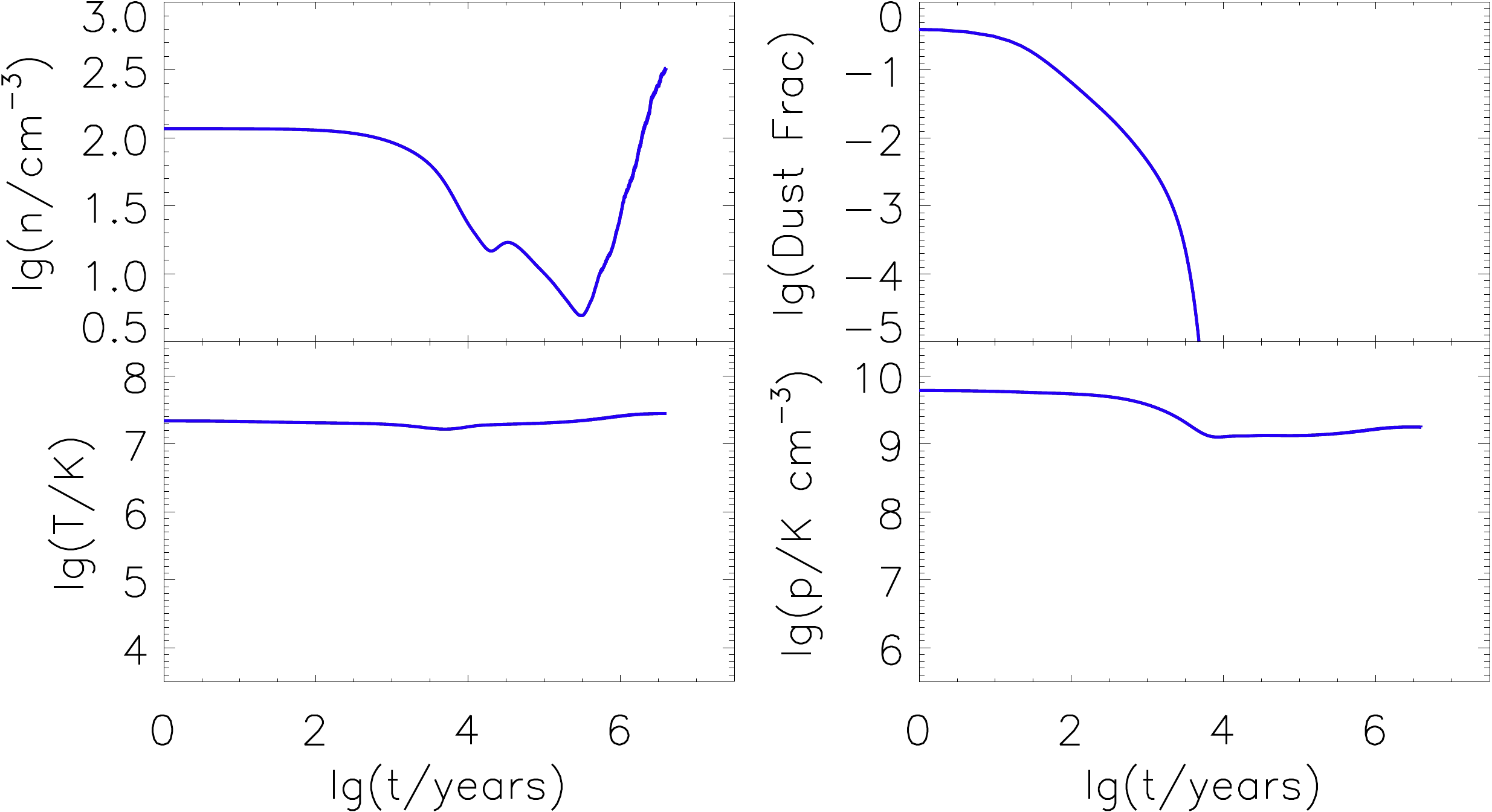}}
\caption{Evolution of thermodynamic quantities in a simulation including electron thermal conduction and the Compton background raised by a factor of 100.  Panels are as in Figure \ref{fig:run1evol}.}
\label{fig:run4evol}
\end{figure*}

\subsection{Enhanced Compton Heating and Conduction}

Finally, we carried out a simulation in which we both artificially boosted the  flux of the Compton background by a factor of 100, and included the impact of electron thermal conduction. In this case, the earliest phases of the evolution follow those of the previous runs, but at later times, electrons carry heat into the cold clumps, disrupting those with size smaller than the Field length (about 25-30 pc).  This disruption of the smallest clumps causes a significant decrease in the RMS density fluctuations, which, by $3 \times 10^5$ years, drop to $\approx 3$ cm$^{-3},$ the smallest value of any of the simulations. This means that only few clumps survive to cool to  8000 K.   

On the other hand, the few clumps that do manage to cool experience strong density enhancements at late times.  This is due not only to the hot, pressurized medium in which they are embedded, but also to the additional compression caused by the pressure wave produced by the conductive evaporation of the clump external layers \citep[e.g.][]{1977ApJ...211..135C, 1977ApJ...215..213M, 2016ApJ...822...31B}.   This leads to rare, but very high density $\geq 10^4$ cm$^{-3}$ clumps, surrounded by a largely uniform hot medium.   Thus while slices from the simulations (not shown) are mostly featureless, the few high density clumps lead to the large RMS fluctuations present at late times in Fig. \ref{fig:run4evol}.  Thus, this case leads to a two phase medium with density contrasts that are even larger than in the case without conduction, although the majority of the mass now remains in the hot phase. 

\begin{figure*}
\center{\includegraphics[width=140mm]{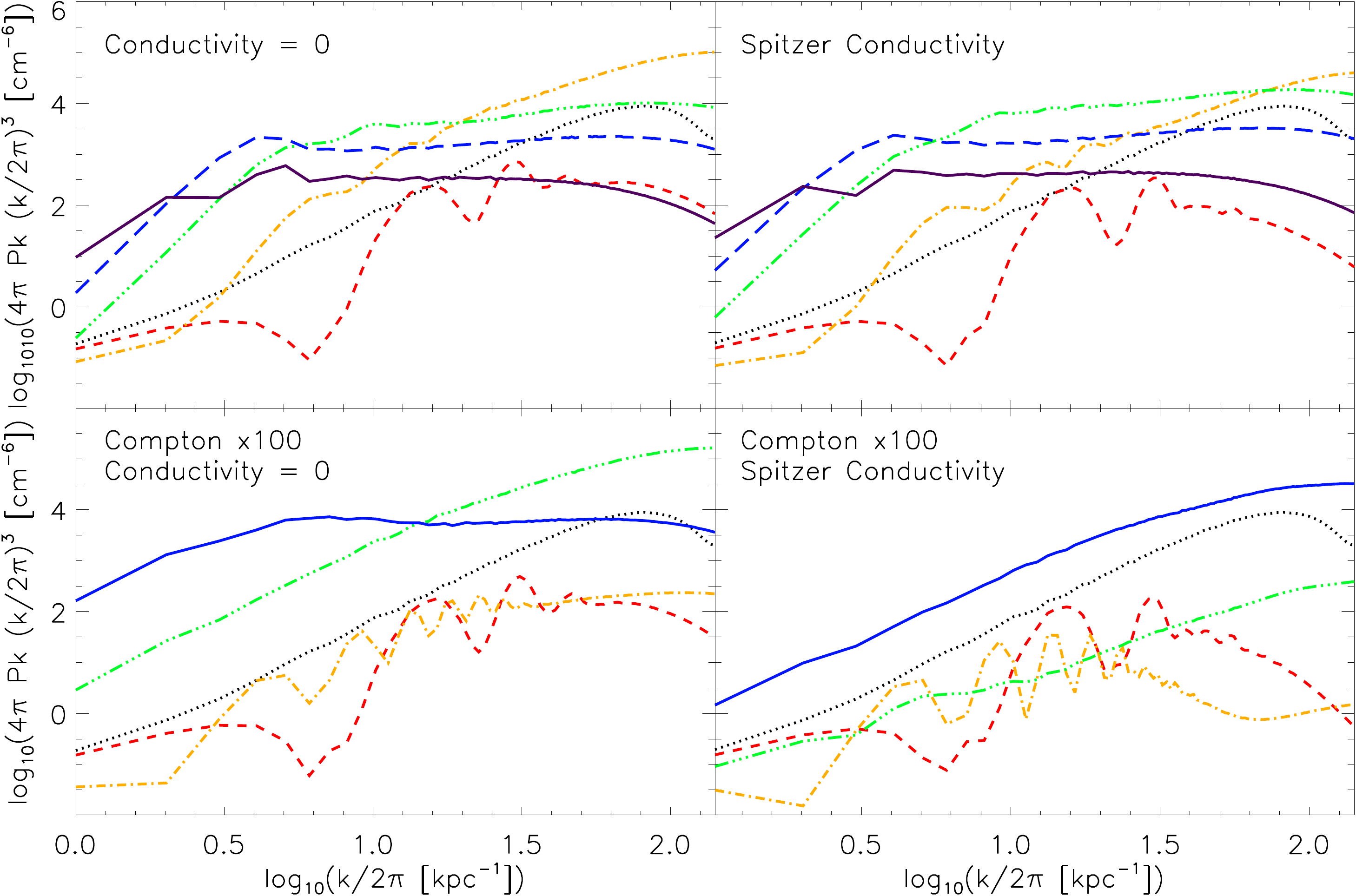}}
\caption{Normalized density power spectra, $4 \pi (k/2 \pi)^3 P(k)$, for our run with $\kappa_e =0$ (top left), our run with $\kappa_e$ given by eq.\ (\ref{eq:thermalcond}, top right),  our run with
$\kappa_e =0$  and the Compton flux boosted by a factor of 100 (bottom left), and our run with $\kappa_e$ given by eq.\ (\ref{eq:thermalcond}) and the Compton flux boosted by a factor of 100 (bottom right).  In all panels the dotted (black), dashed (red), dash-dotted (orange), and dashed trippled-dotted (green) lines represent the configuration at characteristic times of 0 Myr, 0.1 Myr, 0.3 Myr, and 1 Myr, respectively.  In the top panels the long-dashed (blue) and solid (violet) lines show results at 3 and 10 Myrs, while in the lower panels the solid (blue) lines show the configuration at 3 Myrs.}
\label{fig:powespectra}
\end{figure*}

\subsection{Density Distribution}

To quantify the evolution of the structures in our simulations further, we computed the density power spectrum,
\be
P(k) \equiv \left< \tilde n(k) \tilde n^*(k) \right>, 
\ee
where $n({\bf k}) \equiv \int d^3 x \, n({\bf x})  \, \exp(- i {\bf k \cdot x})$ and the wave number $k$ is equal to $2 \pi$ divided by the wavelength of the perturbation.
In this case, the total rms density fluctuations are
\be
\sigma^2 = 4 \pi \int \left(\frac{k}{2 \pi} \right)^3 P(k) {\rm d \, ln k},
\ee
and thus in Fig.\ \ref{fig:powespectra} we show the evolution of the structure in our simulations as quantified by $4 \pi (k/2 \pi)^3 P(k).$ 
The upper panels of this figure quantify the three stages of evolution in the fiducial Compton background case: first initial pressure differences act to rapidly suppress small-scale fluctuations, then larger scale fluctuations are built up as the pressure drops in dense, more rapidly-cooling areas, and finally fluctuations on all scales are damped as the medium as a whole drops to $\approx 8000$K.  The lower panels of Fig.\  \ref{fig:powespectra} correspond to the runs in which the Compton flux has been boosted by a factor of 100.  In this case increased heating in the first stage of the evolution leads to an even more rapid disruption of the the initial fluctuations than seen in fiducial case.  However, at late times,  a lasting two-phase medium develops.   In the $\kappa_e =0$ run this is made up of moderately dense $\approx 10^3$ cm$^{-3}$ structures on a range of scales, but in the conductive case these structures are extremely dense $\gsim 10^4$ cm$^{-3}$ and compressed down to the resolution limit of our simulations.

\section{Summary}
We have studied the origin of the cold molecular clumps recently detected in CO and HCN emission in QSO outflows. We first described the physical properties of the radiation-driven outflow and show that a transition from a momentum- to a energy-driven case must occur at radial distances of $R \approx 0.25$ kpc. During this transition, the dense shell fragments due to Rayleigh-Taylor instabilities, but we find that these clumps are likely to contain very little mass, and they are rapidly ablated and destroyed by the hot gas in which they are immersed.

We have thus also explored an alternative scenario in which clumps form from thermal instabilities at $R \geq 1$ kpc, possibly containing enough dust to catalyze molecule formation. We investigate this processes with 3D two-fluid (gas + dust) numerical simulations of a kpc$^3$ patch of the outflow, including atomic and dust cooling, thermal conduction, dust sputtering, photoionization from the QSO radiation field, and self-shielding (see Appendix B). We find that in all cases dust grains are rapidly destroyed during $\approx 10^4$ years; however, and while cool clumps are present in the fiducial run, they appear only as a transient feature that is washed away as cooling is completed. In fact, a stable two-phase medium with dense clumps is found only if we artificially enhance the QSO radiation field by a factor 100. This result, together with the complete destruction of the dust grains, renders the interpretation of molecular outflows a very challenging problem. 

We pause for a caveat. We have shown that the cooling post-shock gas cannot be piled-up in a well-formed shell due to the fact that QSO outflows, expanding in the steep halo density profile, are Rayleigh-Taylor unstable at all times. In fact, the arguments given for the momentum-driven shell in Sec. 2.2 apply exactly to the energy-driven phase we are discussing. The hot shocked wind gas wraps clumps around and rapidly shreds them.  However, if the gas is thermally unstable, fragments form by thermal instability on a timescale $< 1$ Myr (from our simulations) that is comparable or shorter than the RT growth timescale (see eq. 17, but with $R_C$ now at least 10 times larger). This might lead to a different scenario, that is the one we have explored here. Cold clumps are embedded in a hot interclump medium; the two components are in pressure equilibrium. Hence, clumps are protected from ablation by the hot gas surrounding them, and might survive longer. A more robust conclusion on the clump survival in a accelerated two-phase medium requires additional study. Nevertheless, the above qualitative arguments suggest that this situation is more promising than the RT unstable case.

How the physical conditions leading to clump formation might be realized in the environment of quasars is unclear. Contrary to intuition, cold clumps do not form in the fiducial, standard conditions because the heating is insufficient to support a stable hot phase until the medium as a whole has cooled.   Instead, a stable two phase medium would require a much stronger energy input from the quasar that is inconsistent with the radiated power. One might speculate that other heating mechanisms might be at work. The most likely among these is the contribution of relativistic particles. Although this solution appears unlikely, it is certainly worth further scrutiny, given the thorny questions open by our investigation.  

Our simulations include all the most relevant physics apart from magnetic fields, self-gravity,  and the impact of the bulk motion of the medium as clump formation occurs.  Although anisotropic conduction due to the presence of magnetic fields could slow electron thermal conduction,  conduction not appear to be a key process in determining the final state of the gas.  Furthermore, while gravity  in principle could produce a collapse of the cold clumps, we show in Appendix A, self-gravity can be safely neglected as a result of the outflow expansion, as the gas is like torn apart by expansion on a much shorter time-scale that it can gravitationally collapse.  

This point adds yet another puzzling problem. If self-gravity cannot confine the clumps (in case they form), then they must be pressure confined. However, as the outflow expands to larger radii, the internal density drops and the clumps are progressively less shielded agains the UV quasar radiation acting to dissociate CO molecules. Moreover, if densities become low enough, the cold gas can also become ionized. This might be broadly consistent with the evidence of ionized gas recently found at large distances from the QSO but still clearly associated with it \citep{Nesvadba08, Cresci15, Carniani15}.   

We conclude that the presence of cold molecular clumps in QSO outflows represents a difficult theoretical challenge. Our study shows that neither a scenario in which these components are galactic clouds entrained and accelerated by the outflow, nor the one in which they condense out of the fast, outward moving gas appear to be viable under ``normal'' conditions. Thus, the solution to the problem is in demand of further investigations.

\section*{Acknowledgments} 

We would like to thank C. Feruglio for helpful discussions. This work was supported by National Science Foundation grants AST14-07835 and PHY11-25915 and NASA theory grant NNX15AK82G.  We would like to thank the Texas Advanced Computing Center (TACC) at The University of Texas at Austin, and the Extreme Science and Engineering Discovery Environment (XSEDE) for providing HPC resources via grant TG-AST130021 that have contributed to the results reported within this paper.

 \vspace{0.2in}



%

\vspace{0.2in}

\section*{Appendix A: Multiphase Medium} 

\begin{figure}[t]
\center{\includegraphics[width=115mm]{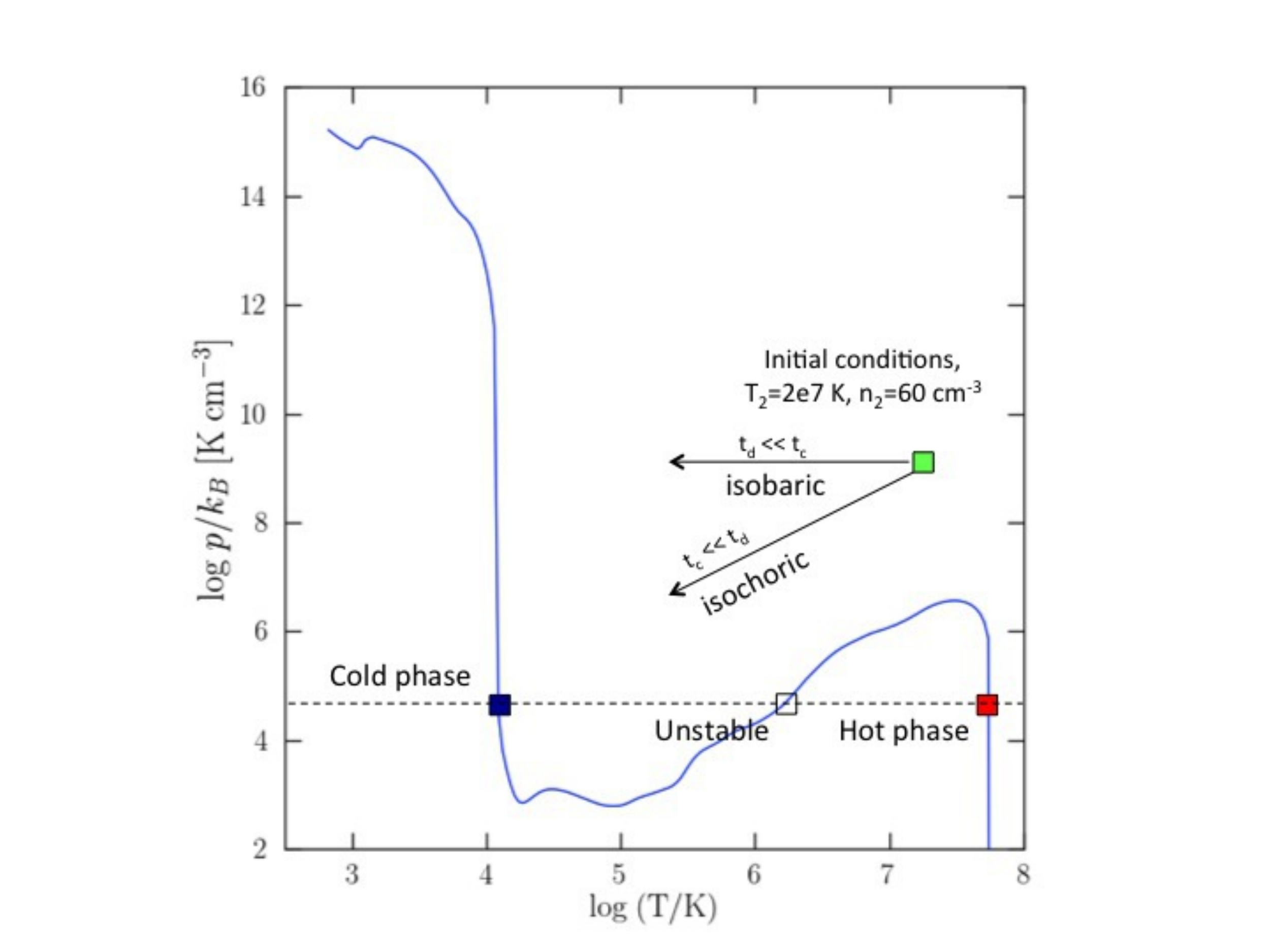}}
\caption{Phase diagram obtained from eq.\ (\ref{eq51}).} 
\label{Fig13}
\end{figure}

Here we show that the hot postshock cooling gas must eventually settle onto a two-phase thermodynamical equilibrium made of a cold phase embedded into a more rarefied hot intercloud medium. 
There are two contributions to the gas heating: photoionization, effective at temperatures $T < T_{eq} \approx 1.9\times 10^5$ K (see eq. \ref{eq30}), and Compton heating arising from the QSO radiation field at temperatures below the Compton temperature, $T < T_C$. If we impose that the gas is thermal equilibrium, i.e. $H= H_\pi + H_C = C$, where $H_\pi$ and $H_C$ are is the photoionization and Compton heating, respectively. Then
\be
\Gamma_\pi \langle h\nu\rangle_\sigma n_H + \Gamma_C \Delta\epsilon\, n_e = n_e n \Lambda(T) 
\label{eq35} 
\ee
or, using the definition of the ionization fraction $x_e$,
\be
\Gamma_\pi \langle h\nu\rangle_\sigma \frac{(1-x_e)}{x_e} + \Gamma_C \Delta\epsilon = n\Lambda(T).
\label{eq36} 
\ee
By further introducing $p= n k_B T$, and defining $p_\pi = \Gamma_\pi \langle h\nu\rangle_\sigma k_B T/\Lambda(T)$ and $p_C = \Gamma_C \Delta\epsilon\, k_b T/\Lambda(T)$, the previous equation is simply written as
\be
x_e = \frac{p_\pi}{p+ p_C+p_\pi}.
\label{eq37} 
\ee
We finally introduce the non-dimensional variables $\Theta = p/p_\pi$, and $\Upsilon = p_C/p_\pi$ to obtain.
\be
x_e = \frac{1}{1+ \Theta + \Upsilon}.
\label{eq49} 
\ee
Recall that the ionization fraction is obtained by imposing photoionization equilibrium:
\be
x_e = \frac{1}{2} \Xi \left[-1+\sqrt {1+ \frac{4}{\Xi}} \right],
\label{eq50} 
\ee
where $\Xi = \Gamma_\pi/n \alpha_B$. Note that for $\Xi \gg 1$, corresponding to heating time scales much shorter than the recombination time in the gas, $x_H = (1-x_e) = \Xi^{-1}$.
By combining eqs.\ (\ref{eq49})-(\ref{eq50}), we obtain an implicit form for the equilibrium curve $H=C$
\be
\Xi \Theta^2 + \Theta \Upsilon^2 + \Xi(\Theta + \Upsilon) + 2\Xi \Theta \Upsilon - 1 = 0. 
\label{eq51} 
\ee

The properties of a multiphase medium are canonically studied in the pressure - temperature phase plane (i.e. the phase diagram) in which the thermal equilibrium curve $H - C = 0$ is plotted.  Eq. (\ref{eq51}) gives the required relation between pressure and temperature. The result is shown in Fig. \ref{Fig13}, where for illustration purposes only we have used a CIE cooling function.  Note that in the numerical simulation we properly account for photoionization effects which, however, do not change the graph qualitatively. From the Figure, we clearly see the existence of a cold ($T \approx 10^4$ K) and a hot phase ($T\approx T_C$) in an extended range of pressures $10^3 < p/k_B < 10^6$ in c.g.s. units. Cooling the gas below $10^4$ K and at the same time keeping thermal equilibrium would require implausibly high pressures. 

One might wonder if the self-gravity might produce a collapse of the cold clumps. The answer is negative, and self-gravity can be safely neglected as a result of the very fast outflow expansion. To see this, let us define the free-fall time as $t_{ff}  \sim (G \rho)^{-1/2}$ and the expansion time scale as $t_{ex} \sim (\dot R/R)^{-1}$. For a spherical expansion, $\rho = \rho_0(R_0/R)^3 = \rho_0 r^{-3}$, where $\rho_0 = \rho_2$ is the density at $R_0=1$ kpc providing the initial condition for our simulation. Recall that the outflow velocity is constant (see eq. \ref{eq13}), i.e. $\dot R = v_{se} =$ const. 
Then, by using the subscript 0 to denote quantities at $R = 1$ kpc we find that 
$t_{ex} = t_{ff}$ is achieved at a value $r_q =(t_{ex,0}/t_{ff,0})^2$. As $t_{ex,0}= 1$ Myr and $t_{ff,0}=15$ Myr, then $r_q < 1$ or $R < R_0$. This implies that  the outflow expansion time is always shorter than the free-fall time. Physically this means that gravity perturbations are ``stretched away" by the expanding flow, or stated differently, the density drops due to outflow expansion so rapidly that gravity perturbations cannot grow. This argument also justifies the neglect of self-gravity in our simulations.

\vspace{0.1in}

\section*{Appendix B: Self-shielding} 
As the gas cools due to thermal instability and condenses into dense clumps which might eventually become optically thick and therefore self-shield against ionizing radiation coming from the quasar. In this case, the photoionization rate, and hence the heating, is considerably reduced and may become very small. A precise treatment of this effect would required radiative transfer calculations that are beyond the scope of the present paper. However, in order to include self-shielding in an approximate manner into the simulations we proceed as follows. 

First we define a self-shielding gas density, $n_S$, such that the opacity to ionizing photons of the gas in a cell becomes unity:
\be
\tau = n_H \sigma_H \Delta x = (1-x_e) n_S \sigma_H \Delta x = 1.
\label{eq41} 
\ee
In the previous equation, $\sigma_H = 6.3\times 10^{-18}$ cm$^{2}$ is the \HI photoionization cross section at 1 Ryd, corresponding to photons with the shortest mean free-path and contributing most to photo-heating; $\Delta x$ is 10 pc, approximately the Field length at the mean initial density. The value of $x_e$ is given by eq. (\ref{eq37}). It follows that
\be
n_S^{-1} = (1-x_e) \sigma_H \Delta x; 
\label{eq42} 
\ee

To catch the photoionization rate reduction, we use a fit to detailed radiative transfer simulation performed by \citet{Rahmati13}. Such formula giver the actual self-shielded photoionization rate, $\Gamma_S$ received by the cell with respect to the optically thin value, $\Gamma$: 
\be
\frac{\Gamma_S}{\Gamma}= 0.98 \left[ 1+ \left(\frac{n}{n_S}\right)^{1.64} \right]^{-2.28} + 0.02 \left[ 1+ \left(\frac{n}{n_S}\right) \right]^{-0.84} 
\label{eq43}.
\ee

\label{lastpage} 
\end{document}